\documentclass[]{emulateapj}

\usepackage{graphicx}
\usepackage{epstopdf}

\newcommand{\is}{\ensuremath{\!=\!}}
\newcommand{\CaII}{Ca~II}
\newcommand{\CaIIH}{Ca~II~H}

\newcommand{\CaIIIR}{Ca~II~854.2~nm}
\newcommand{\Halpha}{H\ensuremath{\alpha}}
\newcommand{\vhor}{\ensuremath{v_{\mathrm{apparent}}}}

\begin{document}

\title{On-disk counterparts of type II spicules in the Ca~II~854.2~nm and
  \Halpha\ lines}

\author{L. Rouppe van der Voort$^1$}\email{rouppe@astro.uio.no} % {v.d.v.l.rouppe@astro.uio.no}
\author{J. Leenaarts$^{1,2}$}\email{jorritl@astro.uio.no}
\author{B. de Pontieu$^3$}\email{bdp@lmsal.com}
\author{M.Carlsson$^{1,2}$}\email{mats.carlsson@astro.uio.no}
\author{G.Vissers$^1$}\email{g.j.m.vissers@astro.uio.no}

\affil{$^1$ Institute of
  Theoretical Astrophysics, University of Oslo, P.O. Box 1029
  Blindern, N-0315 Oslo, Norway}
\affil{$^2$ Center of Mathematics for Applications,
  University of Oslo, P.O. Box 1053
  Blindern, N-0316 Oslo, Norway}
\affil{$^3$ Lockheed Martin Solar \& Astrophysics Lab, Org.\ ADBS,
  Bldg.\ 252, 3251 Hanover Street Palo Alto, CA~94304 USA}

\begin{abstract}
Recently a second type of spicules was discovered at the solar limb with the 
Solar Optical Telescope (SOT) onboard the Japanese Hinode spacecraft. These 
previously unrecognized type II spicules are thin chromospheric jets that are 
shorter-lived (10--60 s) and that show much higher apparent upward velocities 
(of order 50--100 km s$^{-1}$) than the classical spicules. Since they have 
been implicated in providing hot plasma to coronal loops, their formation, 
evolution and properties are important ingredients for a better understanding 
of the mass and energy balance of the low solar atmosphere. Here we report on
 the discovery of the disk counterparts of type II spicules using spectral
 imaging data in the Ca~II~854.2 nm and \Halpha\ lines with the CRisp Imaging 
SpectroPolarimeter (CRISP) at the Swedish Solar Telescope (SST) in La Palma. 
We find rapid blueward excursions in the line profiles of both chromospheric
 lines that correspond to thin, jet-like features that show apparent velocities
 of order 50~km~s$^{-1}$. These blueward excursions seem to form a separate 
absorbing component with Doppler shifts of order 20 and 50 km s$^{-1}$ for the
 Ca~II~854.2 nm and \Halpha\ line respectively. We show that the appearance, lifetimes,
 longitudinal and transverse velocities and occurrence rate of these rapid blue
 excursions on the disk are very similar to those of the type II spicules at
 the limb. A detailed study of the spectral line profiles in these events suggests
 that plasma is accelerated along the jet, and plasma is being heated throughout 
the short lifetime of the event.
\end{abstract}


\section{Introduction} 

Spicules are slender features protruding from the solar limb when
observed in for example the \CaIIH\ and \Halpha\ lines. Classically,
 spicules have been reported to show
velocities of order 20--30 km s$^{-1}$, lifetimes of order 5-10 minutes 
and heights up to 5,000--10,000 km above the photosphere (see
\citet{1968SoPh....3..367B} % Beckers spicule review
and
\citet{2000SoPh..196...79S} % sterling spicule review
for reviews on the older literature on spicules). 

The high spatio-temporal resolution of the Solar Optical Telescope 
\citep[SOT,][]{2008SoPh..249..167T} onboard Hinode \citep{2007SoPh..243....3K}
 have revolutionized our view of spicules.
\citet{2007PASJ...59S.655D} %de pontieu et al, a tale of 2 spicules
showed that spicules can be grouped in two categories based on Hinode/SOT
\CaIIH\ observations of the solar limb. Type~I spicules appear to rise
 up from the limb and fall back again.
These structures show a similar dynamical evolution as active region
dynamic fibrils
\citep{2006ApJ...647L..73H,  % Hansteen et al. DFs
  2007ApJ...655..624D}  % de Pontieu et al. DFs
and a subset of quiet sun mottles
\citep{2007ApJ...660L.169R}, % Rouppe et al. mottles
so they can be interpreted as their off-limb counterparts. Numerical
 simulations based on radiative MHD indicate that
the motion of type~I spicules is driven by magneto-acoustic shocks
 that can be generated by a variety of processes, such
as convective buffeting, granular collapse, dissipation of magnetic 
energy, and leakage of waves into the chromosphere
 \citep[e.g.,][]{2006ApJ...647L..73H,  % Hansteen et al. DFs
  2007ApJ...655..624D, % De Pontieu et al. DFs
  2007ApJ...666.1277H, % Heggland et al. sims
  2009arXiv0906.4446M}. % Martinez-Sykora et al. sims

Type~II spicules show different behavior. They
exhibit upward motion, followed by rapid fading from the Hinode
\CaIIH\ passband, without a downward moving phase. Sometimes they
accelerate while they rise. They have lifetimes between 10 and 100~s,
apparent velocities between 50 and 150~km~s$^{-1}$ and widths between 150 and
700~km and undergo a swaying motion caused by the upward propagation
of Alfv\'enic waves
\citep{2007Sci...318.1574D}. % de Pontieu et al. Alfven waves
These spicules are longest in coronal holes, with heights up to 10~Mm,
while they appear shorter in active regions, where they rarely extend
more than 2~Mm in height. 

Many questions about type~II spicules remain. For example, it is unclear
 whether the apparent motions that are measured at the limb (which are 
based on the temporal evolution of \CaIIH\ intensity in the plane of the sky)
 are asssociated with bulk mass motion. Previous studies have found plentiful 
evidence for line of sight motions of order 20--30 km s$^{-1}$, but little 
evidence for flows of order 50--100 km s$^{-1}$.
Clearly, measuring Doppler shifts on the disk could shed light on this issue.
 However, while type~II spicules are 
ubiquitous at the limb, both in coronal holes and in quiet sun regions
\citep{2007PASJ...59S.655D},  %de pontieu et al, a tale of 2 spicules
it has been unclear what their counterparts are on the solar disk. Finding such 
a disk counterpart is crucial for several other reasons. It can help reveal what
 the formation mechanism of these jets is (which is difficult to study at the limb
 because of the line of sight superpositon). Type~II spicules have also been 
implicated in providing the corona with hot plasma by \citet{2009arXiv0906.5434D} who
established a correlation between faint upflowing signals at coronal temperatures
 (from asymmetries in EUV spectral lines) and the highly dynamic signals in the
 Hinode/SOT \CaIIH\ passband. The latter signals are thought to be associated 
with type~II spicules, but unequivocally establishing the disk counterpart of type~II spicules 
would help reduce the uncertainties involved in establishing the importance of
type~II spicules in the mass and energy balance of the corona. 

In this paper we will focus on establishing what type~II spicules look like 
on the disk.  This is not a straightforward task, given the high spatial and temporal 
resolution required to observe these features.
\citet{1998SoPh..183...91W} % Wang et al. Comparison of prominences in
                            % Halpha and He II 304
%
and
\citet{1998ApJ...504L.123C} % Chae et al. chromospheric upflow events
looked for \Halpha\ jets on the disk with a spatial resolution of
1~arcsec. They found roundish darkenings in the blue wing of the line,
with typical sizes of 3--5~arcsec and lifetimes of 2~minutes, without
subsequent redshift. These darkening are often associated with
converging magnetic dipoles at supergranular boundaries in the
photosphere, even though they also observe some jets in unipolar
regions. Using cloud modeling they find typical upflow velocities of
30~km~s$^{-1}$. While some properties fit with the observed behavior of
off-limb \CaIIH\ type~II spicules, their sizes and lifetimes indicate
that these are not the same phenomenon.

More recently, 
\citet{2008ApJ...679L.167L} % langangen et al IBIS RBEs
investigated so called ``rapid blueshifted events'' (RBEs) in on-disk
\CaIIIR\ data obtained by the Interferometric BIdimensional
Spectrometer (IBIS). RBEs are a sudden widening of the line profile on
the blue side of the line, without an associated redshift. The RBEs
are located around the network -- but not directly on top of
individual network elements, show blueshifts of 15--20~km~s$^{-1}$,
and have an average lifetime of 45~s. The authors interpret these RBEs
as chromospheric upflow events without subsequent downflow and suggest
that these might be the on-disk counterparts of Type~II
spicules. Using Monte Carlo simulations they show that the low
observed blueshift can be explained by a wide range of spicule
orientations combined with a lack of opacity in the upper
chromosphere.

Full 3D non-LTE radiative transfer calculations on a model atmosphere
based on a snapshot of radiation-MHD simulation by
\citet{2009ApJ...694L.128L} % Leenaarts et al 3D 8542 transfer
show that darkenings in the blue wing of the \CaIIIR\ line indeed
correspond to chromospheric upflows.

The evidence for the connection between RBEs and type II spicules
presented by 
\citet{2008ApJ...679L.167L} % langangen et al IBIS RBEs
is tantalizing but not conclusive. 
With their fast evolution and small spatial dimensions, type II
spicules are an elusive phenomenon that are on the limit of what is
observable with present day telescopes. 
In this study we improve on the 
\citet{2008ApJ...679L.167L} % langangen et al IBIS RBEs
observations in terms of spatial and temporal resolution, and employ a
spectral sampling that is better suited to cover high Doppler
shifts. In addition, we observed both the \CaIIIR\ and the \Halpha\
spectral lines. 

\section{Observations and data reduction}

\begin{figure}
  \includegraphics[width=\columnwidth]{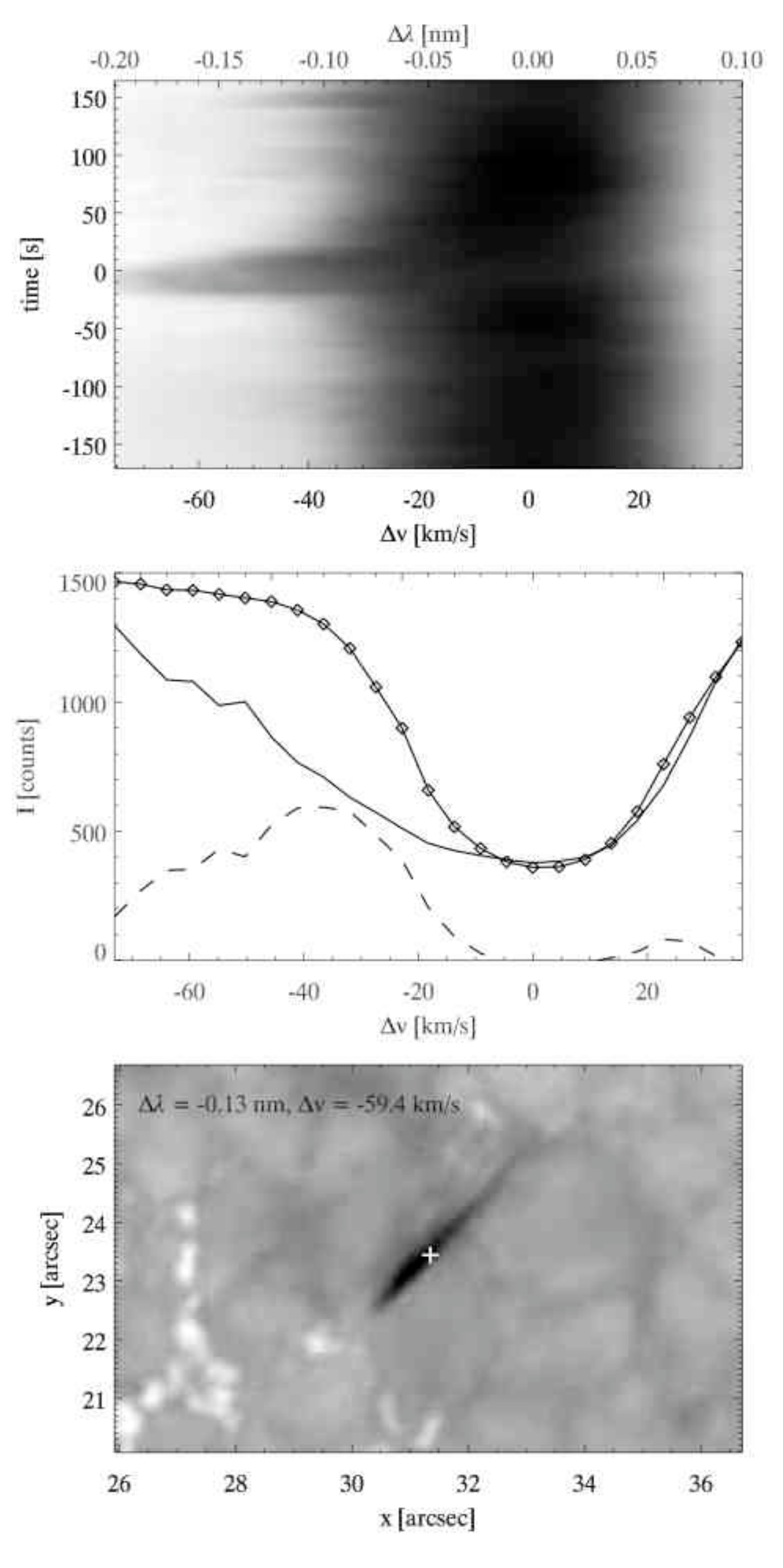}
  \caption{Details from one RBE observed in \Halpha. The top panel
    shows the spectral evolution ($\lambda t$-dataslice) in one
    spatial location in the RBE structure. The location is marked with
    a cross in the \Halpha\ blue-wing image ($\Delta\,v=-59.4$
    km\,s$^{-1}$) shown in the bottom panel. A detailed line profile
    is shown in the middle panel. The profile with open diamonds is
    the \Halpha\ profile averaged over the whole observed FOV, with
    the diamonds marking the sampling positions. The dashed line is
    the difference of the average and RBE profiles. }
    \label{fig:ha-lamt}
\end{figure}

\begin{figure*}
 \includegraphics[width=\textwidth]{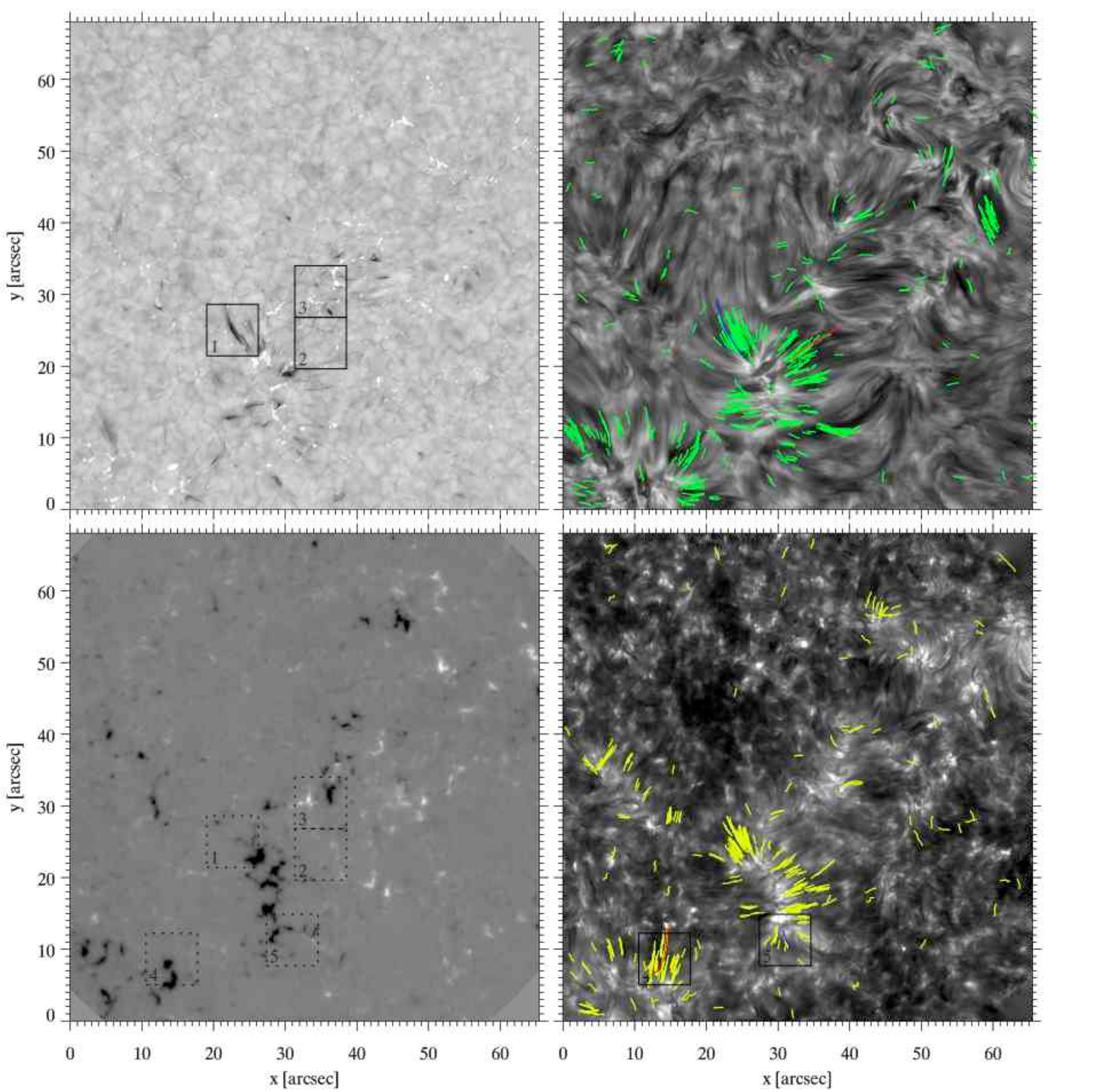}
 \caption{Images from the dataset of 2008 June 15. Top left: blue
   wing of \Halpha\ line at $\Delta v \is -45$ km~s$^{-1}$; Top right:
   \Halpha\ line center; bottom left: Fe 630.2~nm blue wing ($-4.8$ pm) Stokes V/I$_\textrm{cont}$ magnetogram, linearly scaled between $\pm7\%$; bottom
   right: Ca II 854.2~nm line core.  The black boxes indicate regions
   of interest (ROIs) shown in Figs.~\ref{fig:rbe-cuts-ha},
   \ref{fig:rbe-cuts-bp}~and~\ref{fig:rbe-cuts-ca}. The yellow  and green
   lines show detected paths of respectively \CaIIIR\ and \Halpha\ RBEs. 
   The RBEs along the red and blue lines
   are shown in detail as images in Figs.~\ref{fig:rbe-cuts-ha}
   and~\ref{fig:rbe-cuts-ca} and as time-slices along their
   trajectories in Figs.~\ref{fig:xt-slice-ha}
   and~\ref{fig:xt-slice-ca}. 
%The top-left panel shows an RBE as a
%  diagonal dark streak at $(x,y) \is (22,24)$ and two small roundish
% ''black beads'' at $(x,y) \is (35,27)$ and $(34,30)$. 
   Note that the \Halpha\ line core image shows significantly more
   fibrilar structures than the \CaIIIR\ image, indicating that the
   \Halpha\ line has the largest chromospheric opacity of the
   two. \label{fig:fov}}
\end{figure*}

The observations were obtained with the CRisp Imaging
SpectroPolarimeter 
\citep[CRISP,][]{2008ApJ...689L..69S} % scharmer et al. CRISP
installed at the Swedish 1-m Solar Telescope
\citep[SST,][]{2003SPIE.4853..341S} % scharmer et al. SST
on La Palma (Spain).
CRISP is a spectropolarimeter that includes a dual Fabry-P{\'e}rot
interferometer (FPI) system similar to that described by
\citet{2006A&A...447.1111S}. % Scharmer: Fabry-Perot design
CRISP is equipped with 3 high-speed low-noise CCD cameras that operate
at a frame rate of 35 frames per second and an exposure time of 17~ms.
The 3 cameras are synchronized by means of an optical chopper, 2 of
these cameras are positioned behind the FPI after a polarizing beam
splitter, and the 3rd camera is positioned before the FPI but after
the CRISP pre-filter. The latter camera is used as anchor channel for
image processing and is referred to as the wide-band channel.
The image scale is 0\farcs071/pixel and the field of view (FOV)
$\sim$71$\times$71 arcseconds.
CRISP allows for fast wavelength tuning ($\lesssim$50~ms) within a
spectral region and is ideally suited for spectroscopic imaging of the
chromosphere where the dynamical evolution time can be on the order of
a few seconds, sometimes even faster than 1~s
\citep{2006ApJ...648L..67V}.
For \Halpha\, the transmission FWHM of CRISP is 6.6~pm, and the
pre-filter FWHM is 0.49~nm. 
For \CaIIIR\, the transmission FWHM is 11.1~pm, and the pre-filter
FWHM is 0.93~nm.
%

% ApJ style rules: Dates are written in the order: year, month, and
% day; e.g., 1996 January 1. In tables, use three-letter abbreviations
% for months, without a period. Universal time designations are
% written 22:37:48 - 22:37:52.5 UT (for hours, minutes, seconds). 
In this paper, we analyze two data sets from 2008 June 15, from an area
at the edge of a coronal hole close to disk center at
$(x,y)\approx(150,-28)$ ($\mu=0.99$): a 40~min \CaIIIR\ time sequence
started at 08:24~UT, and a 24~min \Halpha\ sequence started at
09:15~UT.
The following wavelength sequences were used.
For the \CaIIIR\ data set we used a 29 line position
sequence, ranging from $-$200~pm to $+$200~pm from line center
($\pm$68~km~s$^{-1}$), with 10~pm steps in the core and 20~pm steps
in the wings. 12 exposures per wavelength position were recorded,
resulting in a time of 11~s to complete a full wavelength scan.
For the \Halpha\ data set we used a 25 line position
sequence, with 10~pm steps ranging from $-$160~pm to $+$80~pm from line
center ($-$73 to $+$37~km~s$^{-1}$ Doppler shift). At
each wavelength position, 8 exposures were recorded. The time to
complete a full wavelength scan was 6.7~s. 

The image quality of the time sequences benefited from the SST
adaptive optics system 
\citep{2003SPIE.4853..370S} % Scharmer et al: SST AO
and the image restoration technique Multi-Object Multi-Frame Blind
Deconvolution \citep[MOMFBD,][]{2005SoPh..228..191V}. 
All images from all wavelength positions in a scan were divided in
overlapping 64$\times$64 pixel subfields and processed as a single
MOMFBD restoration. 
In such a restoration, the wide-band channel served as anchor for the
narrow-band CRISP exposures to ensure precise alignment between the
restored narrow-band images.
We refer to \citet{2008A&A...489..429V} for more details on the MOMFBD
processing strategies on similar extensive multi-wavelength scans.

After MOMFBD reconstruction of the individual line scans, the images
from the different time steps were combined to form time series. 
The images were de-rotated to account for diurnal field rotations, and 
aligned and de-stretched to account for translation and warping due to
seeing effects. 
(Local) offsets were determined on the wide band images and applied to
the corresponding CRISP images. 

The \CaIIIR\ dataset is of excellent quality, with most images close
to the diffraction limit of 0.23~arcsec. 
The \Halpha\ dataset is not as good, as 
the seeing was degrading in quality and 
there were more frequent
short moments of blurring, but still at least half of the images are
close to the diffraction limit of 0.17~arcsec.

We obtained a context Stokes V magnetogram from a CRISP scan of the Fe
630.25~nm line recorded at 09:39 UT, after the \CaIIIR\ and \Halpha\
sequences. 
We followed the procedures of 
\citet{2008ApJ...689L..69S} % scharmer et al. CRISP
except that we did not use MOMFBD restoration on the polarimetric observations. 
In the lower left panel of Fig.~\ref{fig:fov}, a blue wing ($-$4.8~pm) Stokes V magnetogram is shown. 
The magnetogram is the result of flatfielding and adding of in total 32 exposures -- 8 exposures per liquid crystal state. 
We estimate the noise level to be $2.6 \cdot 10^{-3}$ for Stokes V relative to I$_\textrm{continuum}$, and the spatial resolution better than 0\farcs5. % very conservative ... probably ~0.3!
The magnetogram was aligned to the time sequences using the wide-band
channels. As the magnetogram was observed after the time sequences, no
detailed information on the individual magnetic elements can be
obtained but it serves the purpose of giving general information on
the distribution of the different magnetic polarities over the FOV.

We searched both datasets for RBEs. In blue-wing images they appear as
roundish or elongated dark structures on top of the photospherically
formed background
\citep{2006A&A...449.1209L,2006A&A...452L..15L}. %Leenaarts et al BP contrast
In order to efficiently explore the datasets, we used widget based
analysis tools programmed in the Interactive Data Language (IDL): CRIsp
SPectral EXplorer (CRISPEX) and Timeslice ANAlysis Tool (TANAT). 
These tools allow simultaneous display of spectra, images and
$xt$-slices of arbitrarily shaped curves. 
This way it is easy to distinguish RBEs from intergranular lanes, dark
reversed granulation and, by visual inspection of their spectral signature, of
shockwaves.

Our method to find and study RBEs in our data is different from the
\citet{2008ApJ...679L.167L} % langangen et al IBIS RBEs
approach. They identified RBEs purely on a spectroscopic basis,
searching for specific spectral signatures in $\lambda t$-dataslices. 
We find numerous examples of short-lived, narrow, and elongated
features in our data at wavelength offsets corresponding to
significant Doppler velocities.
These features emerge from regions with magnetic field concentrations
and have a similar spectral signature in $\lambda t$-dataslices as the
\citet{2008ApJ...679L.167L} RBEs.

To expand on the statistical properties of the RBEs we augmented the above
described manual detection scheme with an automated algorithm. This algorithm finds 
RBEs by isolating the location and time of long thin features at high blueshifts
 (respectively 30 and 60 km~s$^{-1}$ for the \CaIIIR\ and \Halpha\ datasets). 
To reject intergranular lanes and other non-RBE associated features, we impose 
that the features not be too curved. 
This automated procedure identified 413 features in the \CaIIIR\ dataset and 608 in the
\Halpha\ dataset. We confirmed the validity of the automated search results by 
visually inspecting blue wing images and $\lambda t$-dataslices for the flagged RBEs. 
This inspection confirmed that the automatically detected features have spectral 
characteristics similar to the RBEs of \citet{2008ApJ...679L.167L}. 

In addition to these events, we also (manually) found a number of small, roundish 
features of high blueshift that the algorithm failed to detect (see Sec.~\ref{sec:bb}).

\section{Results}

\begin{figure*}
% \plottwo{figs/2008-06-15_ha_315_348_119.eps}{figs/2008-06-15_ha_486_323_108.eps}
 \plottwo{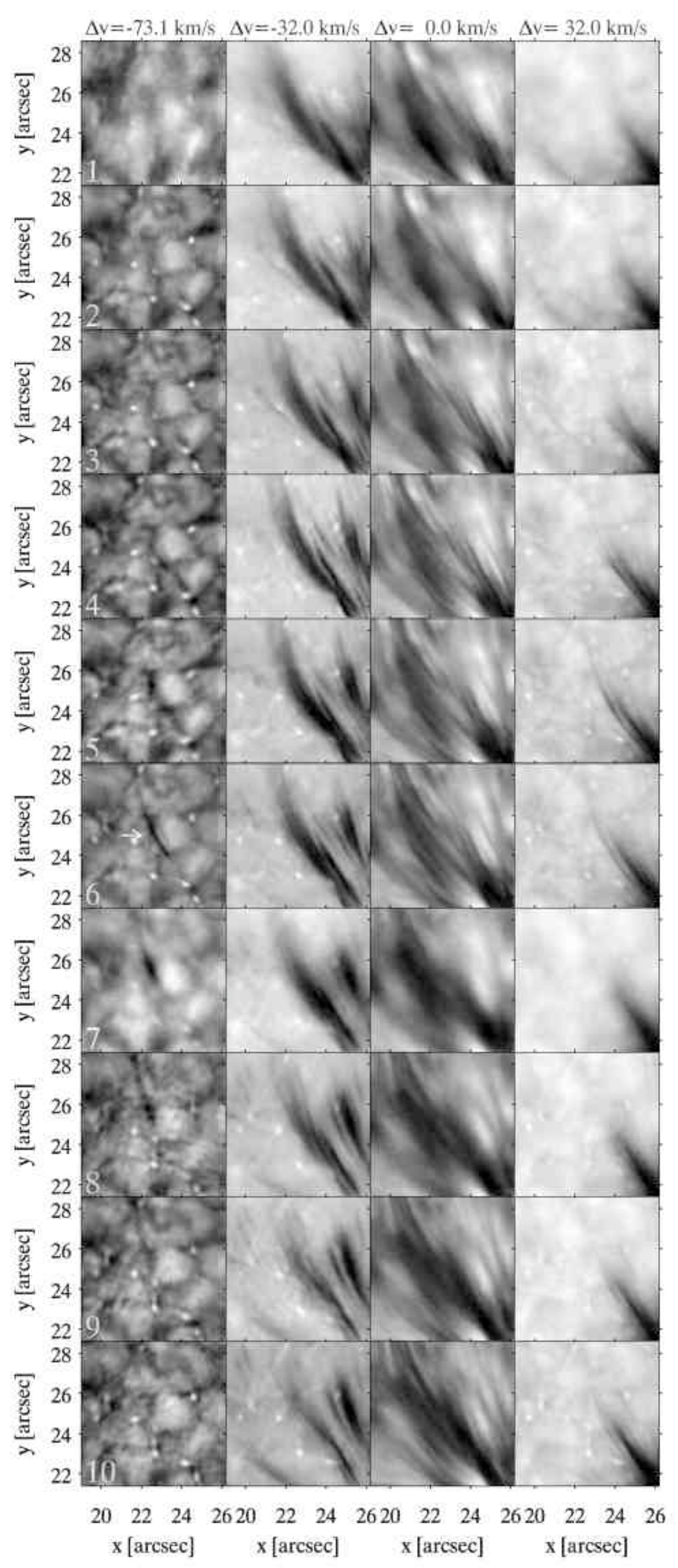}{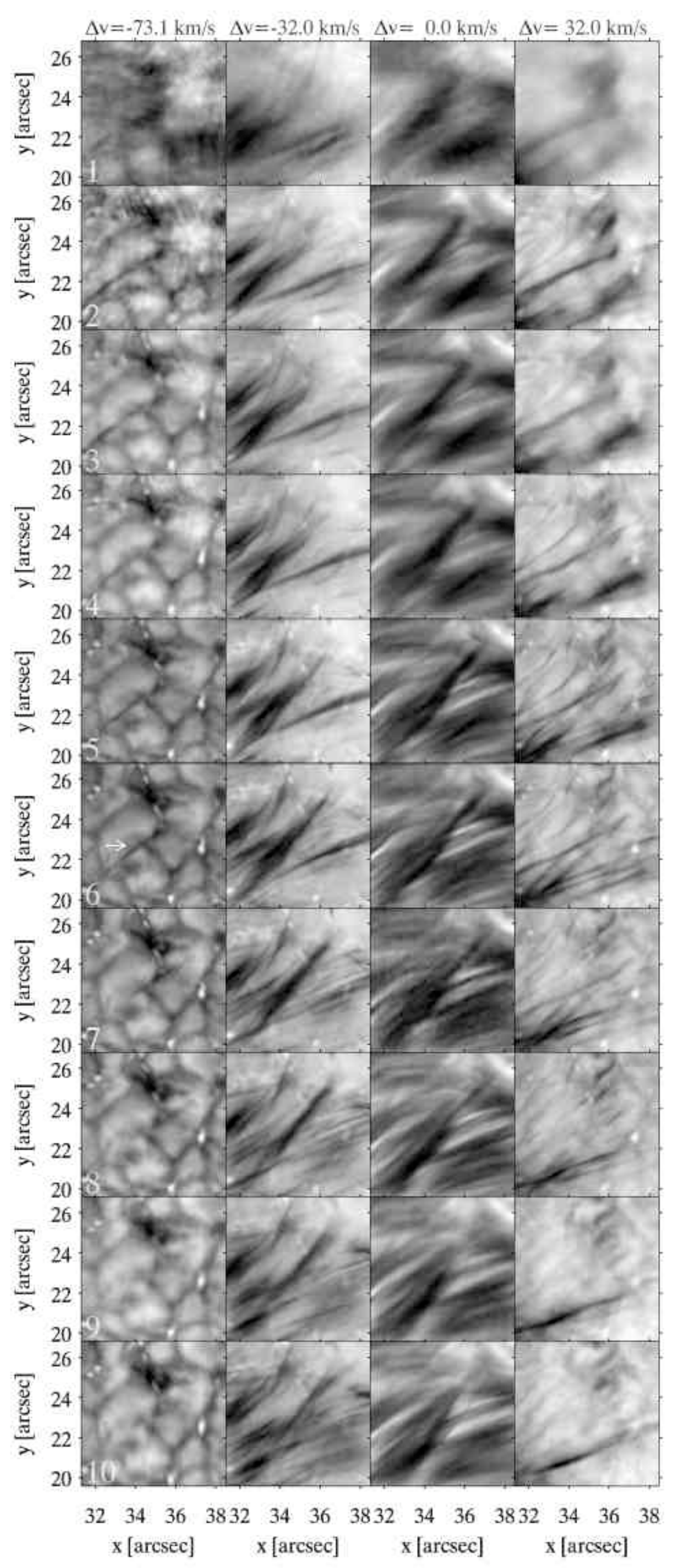}
 \caption{Detailed cutouts of ROI1 (left) and ROI2 (right) showing the
   time-evolution of \Halpha\ RBEs in different line positions as
   indicated above the top row of panels. A negative velocity means a
   blueshift. Time increases downward, with 6.7 s between successive
   images in a column. The RBEs appear as thin dark streaks moving
   through the field of view in the leftmost column at a doppler shift
   of $-$73.1 km~s$^{-1}$. The arrows indicate the RBEs when they are
   the most visible.  They move away from the network, with their
   trajectory roughly aligned with the fibrils visible in the
   line-core images, but no one-to-one correspondence between fibrilar
   structures at the other positions in the line. Time-lapse
   animations of this figure are provided
   electronically. \label{fig:rbe-cuts-ha}}
\end{figure*}

\begin{figure*}
%  \plottwo{figs/halpha_15Jun2008_17Apr2009_094512.csav_t119.eps}{figs/halpha_15Jun2008_17Apr2009_094051.csav_t108.eps}
  \plottwo{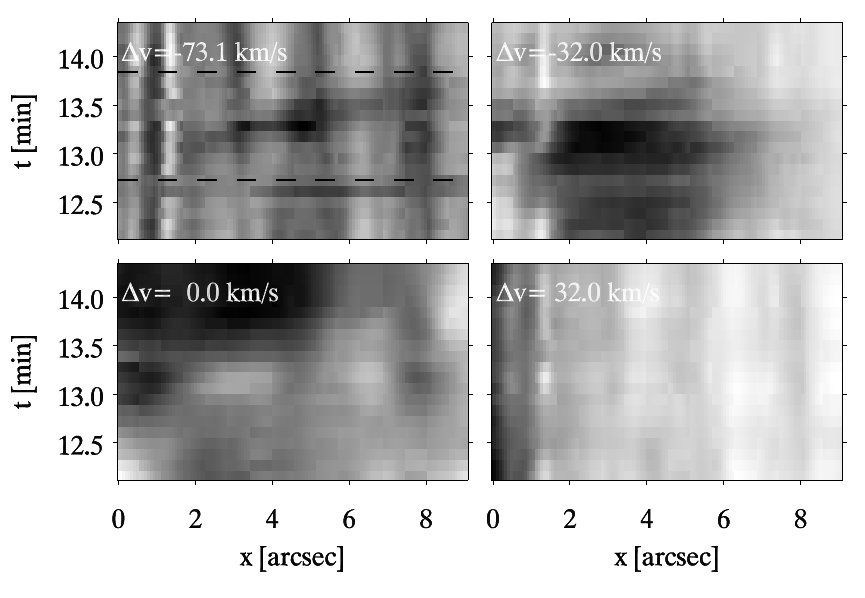}{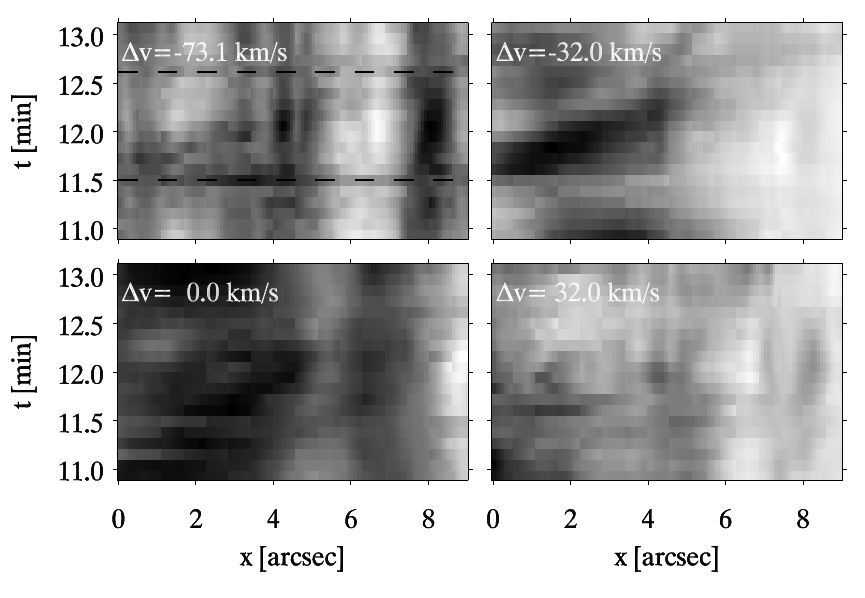}
  \caption{Timeslice along the trajectory of the RBEs in ROI1
    (left-hand panels) and ROI2 (right-hand panels). They appear as
    diagonal streaks in the $-$73.1~km~s$^{-1}$ panels (they move from left to
    right). The $-$32 km~s$^{-1}$ panels show a darkening just before the RBE,
    extending less far away from the network than the RBE itself. The
    black dashed lines indicate the timespan displayed in
    Fig.~\ref{fig:rbe-cuts-ha}. \label{fig:xt-slice-ha}}
\end{figure*}

Figure~\ref{fig:ha-lamt} shows details of one selected RBE observed in
\Halpha. 
The bottom panel shows part of a blue-wing \Halpha\ image at an offset
from line center equivalent to a Doppler shift of
$-$59.4~km\,s$^{-1}$.
The narrow dark streak in the center is identified as an RBE.
% t=174, first occurrence at t=168 -> 6*6.7=40.2 s
% disappeared at t=176 -> total life time=53.6 s
At this wavelength position, the first sign of this RBE appeared 40~s
earlier, close to the photospheric bright point at the lower left of
this feature. 
During its lifetime, the streak moves away from that bright point and
vanishes after 54~s. 
The white cross marks the spatial position where a detailed spectrum
is showed in the middle panel, and the spectral/temporal evolution in
the top panel. 
The detailed spectrum is compared to the average \Halpha\ profile: the
dashed line represents the difference of the two profiles. 
The RBE spectrum displays significant asymmetry towards the blue wing,
differing from the average profile even beyond a Doppler velocity of
72~km\,s$^{-1}$. 

The top panel shows the temporal evolution of the spectra in the
central part of an RBE in a $\lambda t$-dataslice.
This neatly illustrates why these events are called ``rapid blue
excursions'' as they suddenly develop a highly asymmetric line profile
towards the blue.
At this spatial location, there is no spectral signature pointing to
any down-fall phase associated with this event (i.e., features
occurring on the red side of the line).

Figure~\ref{fig:fov} shows the full field-of-view of the datasets of
June 15. % note ApJ date convention
The upper-left panel shows a blue-wing \Halpha\ image at a Dopplershift
of $-$45~km~s$^{-1}$. It shows bright network elements and
many blueshifted chromospheric structures.
At this Dopplershift, one can identify many narrow elongated
features. However, some of these features are not identified as RBEs
as they have different spectral signatures (i.e., not asymmetric to
the blue wing) and/or display different temporal behavior (i.e., no
sudden, or rapid, appearance).

The upper-right panel shows the simultaneous \Halpha\ line-core image.
Green curves indicate the trajectories of the \Halpha\ RBEs we found in an automated fashion -- a dense
concentration was found in the region of interest (ROI) marked `1'.
The RBEs found in ROI3 appear to have different behavior than the
others, with movements perpendicular to the magnetic field structure
as outlined by the fibrils in the line core image.

The lower-left panel shows the context magnetogram. It shows two
regions of unipolar network in the lower left quadrant, and a small
bipolar region at $(x,y) \is (35,35)$~arcsec. The right side of the
field shows scattered patches of unipolar field with opposite polarity
from the two network regions. The lower-right panel shows the \CaIIIR\ line-core, with the \CaIIIR\ RBE
trajectories overplotted in yellow.

The RBEs are predominantly found around the two patches of unipolar
network in the lower-right quadrant of the FOV with a minority of the
RBEs scattered throughout the FOV. The \Halpha-core image
shows much more fibrilar structure than the \CaIIIR\ one. The latter
shows an internetwork acoustic shock pattern
\citep[see e.g.,][]{2009A&A...494..269V} % Vecchio et al. IBIS
                                % acoustic
except very close to the network. This indicates that \Halpha\ has a
much higher chromospheric opacity, and is likely to show structures at
larger height above the solar surface.

\subsection{Rapid blue excursions in \Halpha} \label{sec:rbe-ha}

\begin{figure}
  \includegraphics[width=\columnwidth]{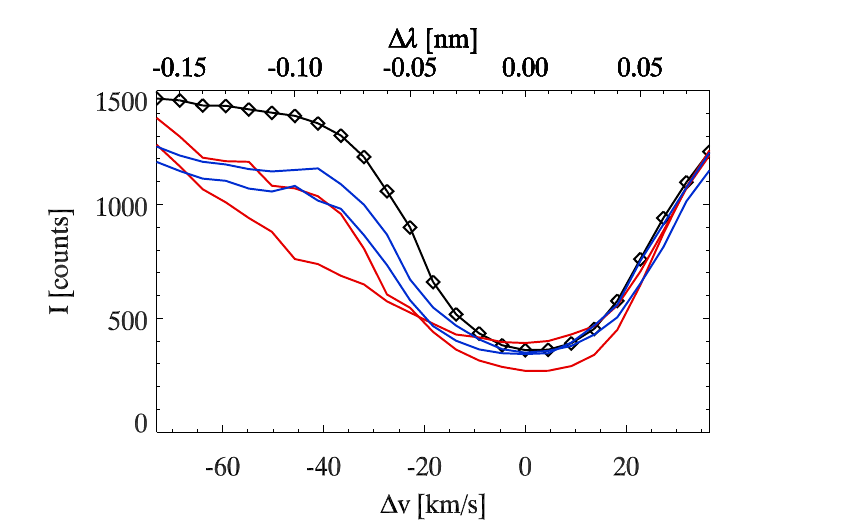}
  \caption{Example line profiles of the \Halpha\ RBEs. Black: average
    spectrum over the field of view, with diamonds indicating the
    different wavelength positions of the filter. Red: profiles of
    RBEs; blue: profiles of ``black beads''.  \label{fig:ha-detsp}}
\end{figure}

Figure~\ref{fig:rbe-cuts-ha} shows blowups of ROI1 and ROI2, following
the detailed evolution of two RBEs whose trajectories are colored red
and blue in the top right panel of Fig.~\ref{fig:fov}. They appear as
thin elongated dark streaks in the $\Delta v \is -73.1$~km~s$^{-1}$
column.

We strongly encourage the reader to view the time-lapse animation of
this Figure that is available on-line. Presenting the temporal
evolution of these narrow, short-lived features on paper as a series
of images cannot rival the visual impression of an animation.

In the left-hand panels the RBE from ROI1 is shown. It appears in the
lower-right corner of frame 2 (numbers in the lower left corner). It
then increases in visibility, becoming darker and longer while moving
diagonally towards the upper-left corner of the ROI. While it
traverses it starts to fade and is almost invisible in frame 10. The
structure is slightly curved and follows a curved path, which suggests
movement along a magnetic field line. 

The left-hand panels of Fig.~\ref{fig:xt-slice-ha} show the time
evolution of the brightness along the trajectory of the RBE. It
appears as a diagonal streak in the $\Delta v \is -73.1$~km~s$^{-1}$
panel. From this panel its apparent horizontal speed can be determined as
$84$~km~s$^{-1}$.

The RBE from ROI2 is shown in the right-hand panels of
Fig.~\ref{fig:rbe-cuts-ha} . It is present from frame 2, sticking out
from the lower-left corner towards the center of the ROI. It does not
appear to move until frame 4. It then moves towards the
upper-right in frames~5--8. It fades and is completely gone in
frame~10. The right-hand panels of Fig.~\ref{fig:xt-slice-ha} show an
$xt$-slice of the brightness along the trajectory of the RBE. There
seems to be a structure with similar horizontal speed in the
$-$32~km~s$^{-1}$ panel (the black structure between $x \is 0$ and $x\is
5$~arcsec). Close inspection of the image sequence in
Fig.~\ref{fig:rbe-cuts-ha} shows that it is caused by a number of
different fibrils and most of the absorption is unrelated to the
RBE. The apparent horizontal speed of the RBE is $114$~km~s$^{-1}$.

The trajectories of both RBEs are roughly aligned with the fibrils and
mottles visible in the wavelengths closer to the line core, but there
is no one-to-one relationship between the RBEs and the dark structures
in the $\Delta v \is -32$~km~s$^{-1}$ indicating that the RBEs are
indeed structures with a large line-of-sight velocity component, and
not the damping wing of an optically thick chromospheric structure
with a small velocity. There is no corresponding absorption in the red
wing during or after the RBE. This rules out that our RBEs are
high-velocity type~I spicules or a related shock phenomenon.

The two red curves in Fig.~\ref{fig:ha-detsp} show typical line
profiles of RBEs. They are strongly asymmetric, with an extended
absorption wing on the blue side of the core relative to the average
profile. The largest excess absorption occurs between Doppler shifts
of $-$60 and $-$30 km~s$^{-1}$.

\subsection{``Black beads'' in \Halpha} \label{sec:bb}

\begin{figure}
  \includegraphics[width=\columnwidth]{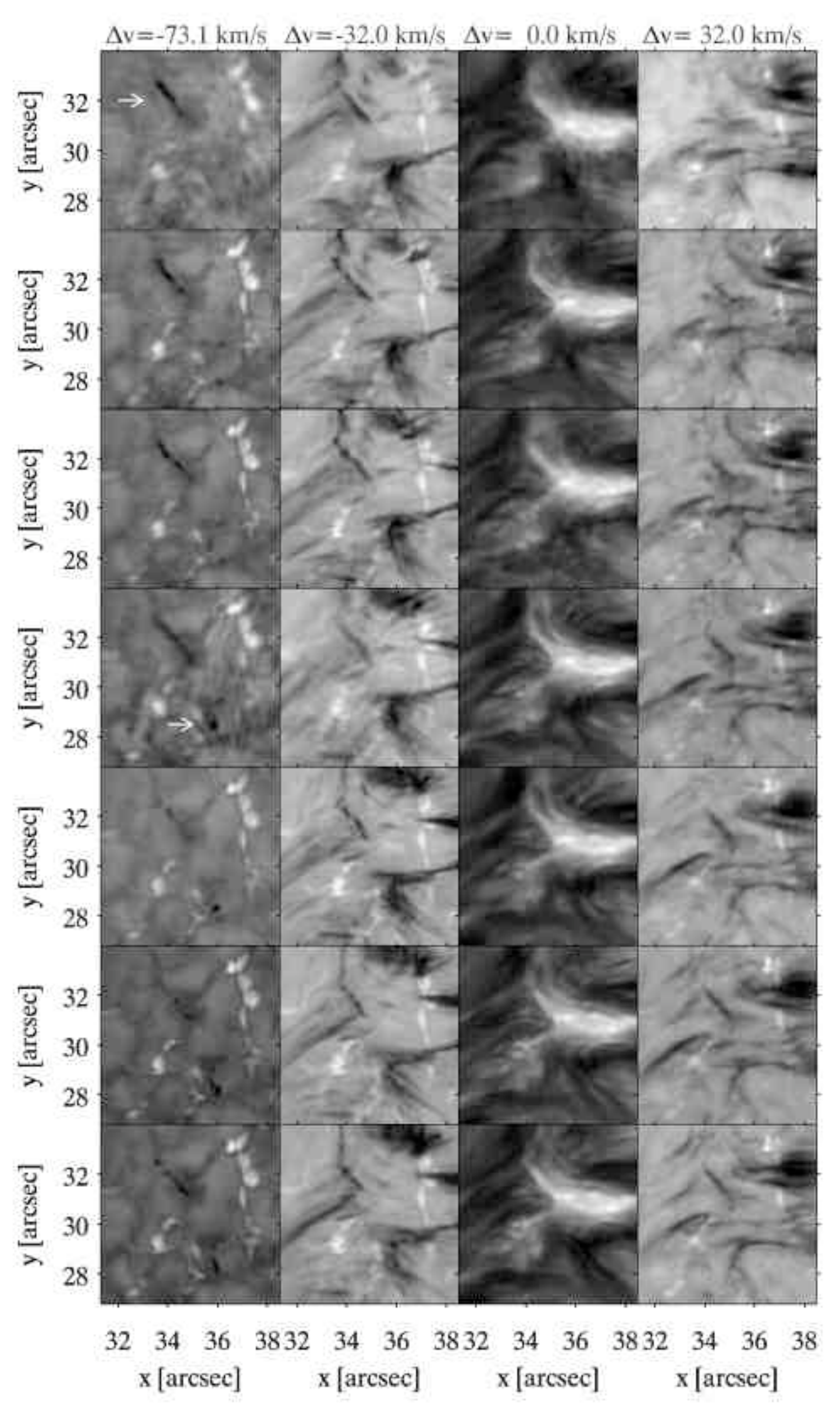}
  \caption{Detailed cutouts of ROI3, showing the time-evolution of
    \Halpha\ ``black beads'' in different line positions as indicated
    above the top row of panels. A negative velocity means a
    blueshift. Time increases downward, with 6.7 s between successive
    images in a column. The white arrows indicate the black
      beads when they are most visible. A time-lapse animation of
    this figure is provided electronically. \label{fig:rbe-cuts-bp}}
\end{figure}

We found a large number of RBEs in ROI3. They are different in several
respects from the RBEs discussed in
Sec.~\ref{sec:rbe-ha}. Figure~\ref{fig:rbe-cuts-bp} shows an image
sequence of ROI3 with several of these special RBEs, which we dub
``black beads''. 
They appear as roundish darkenings, either
individually or in short strings of several beads and are located
along a neutral line between two photospheric magnetic field
concentrations of opposite polarity (see the magnetogram in
Fig.~\ref{fig:fov}). Their horizontal speeds are relatively low 
($15$ km s$^{-1}$) and they move parallel to the neutral line, but
perpendicular to the large scale fibril orientation (see the
\Halpha-core panel of Fig.~\ref{fig:fov}). The small scale magnetic
field configuration close to the neutral line is complex, as indicated
by the appearance of the line-core images. It is unclear how the
horizontal motion of the black beads relates to this configuration.

The line profiles of the black beads (see the blue curves in
Fig.~\ref{fig:ha-detsp}) are even more asymmetric than the line
profiles of the RBEs associated with the unipolar network, with nearly
flat line profiles between~$\Delta v \is -74$ and $\Delta v \is
-50$~km~s$^{-1}$.  They show strong absorption all the way up to the
most blueward position of our line scans, which suggests that the
black beads are structures that, at least partly, move with an upward
velocity well in excess of 70~km~s$^{-1}$.

Typically, these black beads have diameters between 0.15 and 0.3~Mm,
as measured at $\Delta v=-74$~km~s$^{-1}$.

\subsection{Rapid blue excursions in \CaII~IR}

\begin{figure*}
% \plottwo{figs/2008-06-15_ca_197_120_099.eps}{figs/2008-06-15_ca_431_157_119.eps}
 \plottwo{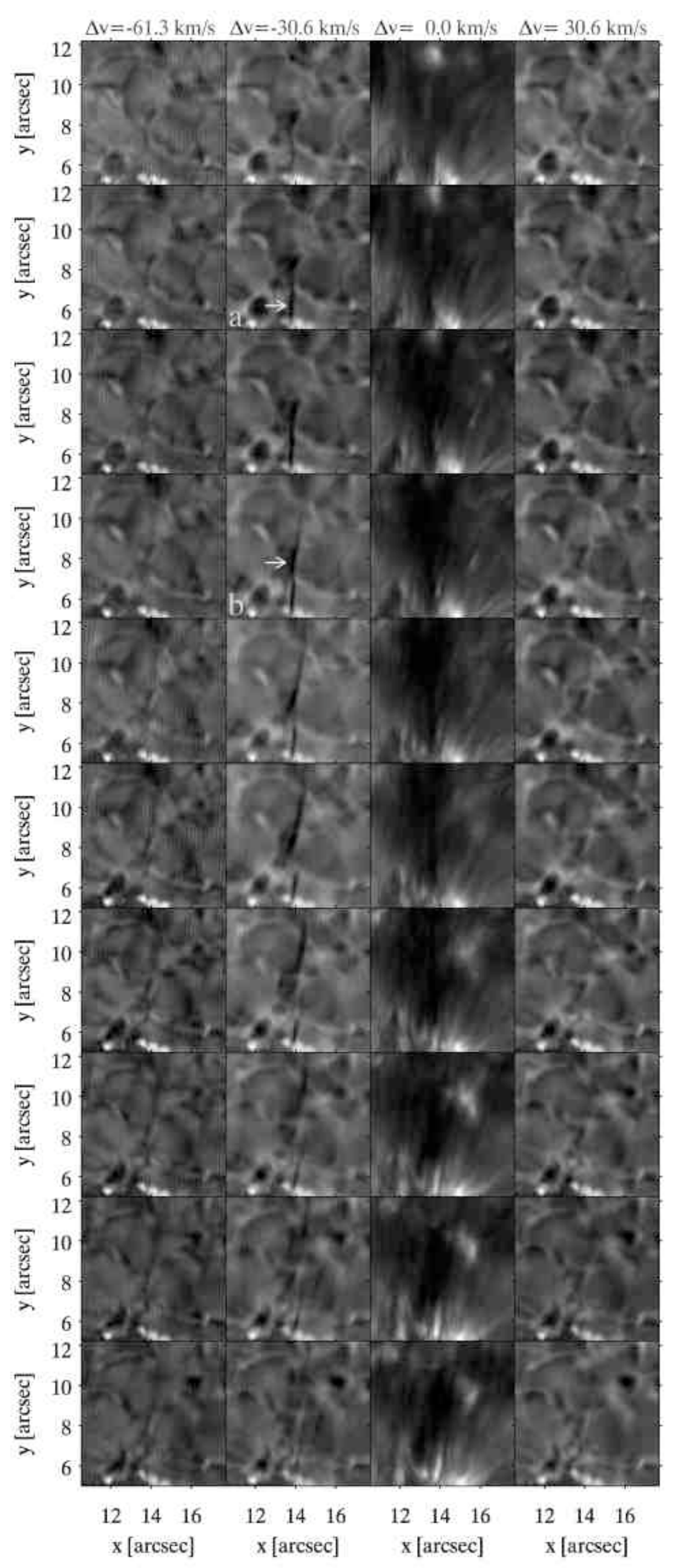}{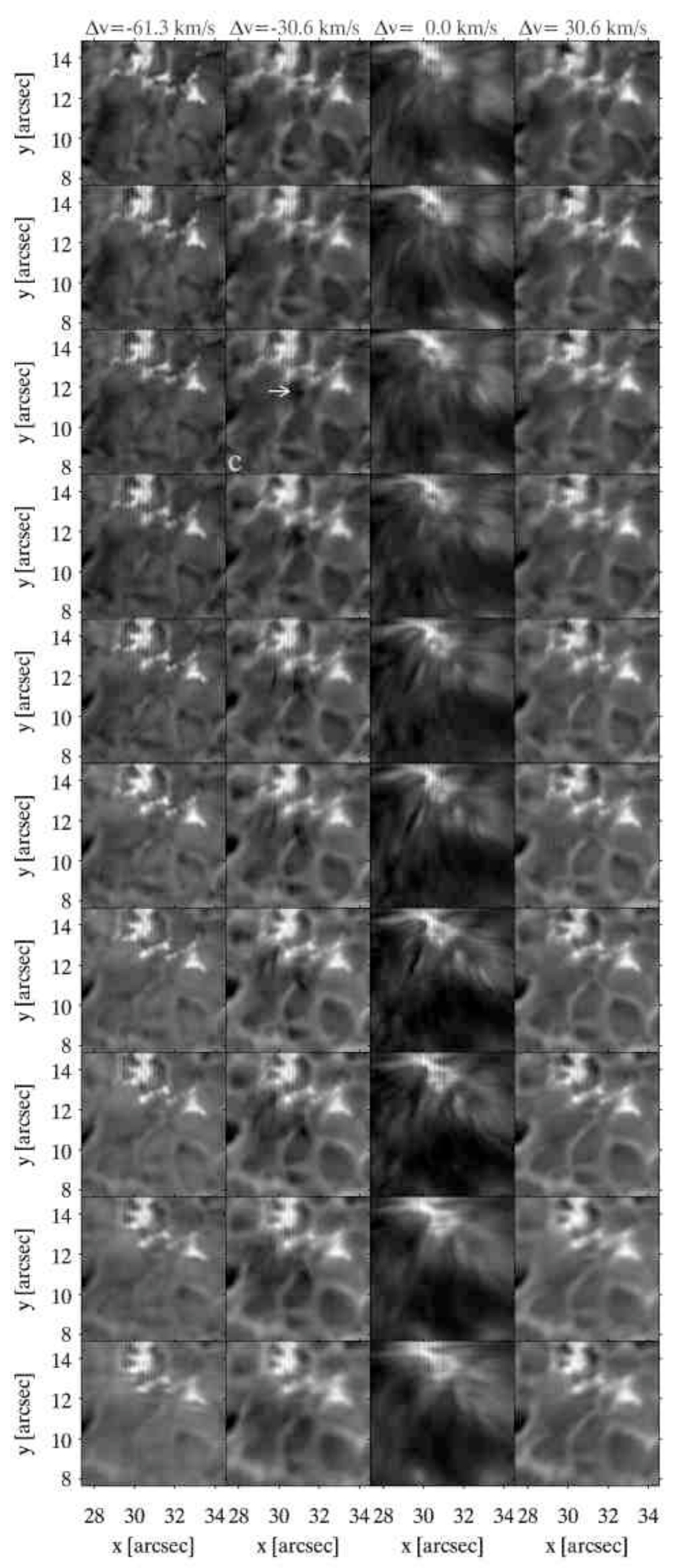}
 \caption{Detailed cutouts of ROI4 (left) and ROI5 (right) showing the
   time-evolution of \CaIIIR\ RBEs in different line positions as
   indicated above the top row of panels. A negative velocity means a
   blueshift. Time increases downward, with 11.35~s between successive
   images in a column. The \CaIIIR\ RBEs are seen in the $-$30.6
   km~s$^{-1}$ row, without a signature in the $-$61.3 km~s$^{-1}$
   column.  The left-hand panels show two RBEs. The first one appears
   in the panel marked `a' from the bright point at $(x,y) \is
   (13.7,5.2)$ and moves straight up. The second one appears in the
   panel marked `b' and is slightly curved. The right-hand panels show
   a slow-moving RBE appearing in the panel marked `c'. Time-lapse
   animations of this figure is provided
   electronically. \label{fig:rbe-cuts-ca}}
\end{figure*}

\begin{figure*}
%  \plottwo{figs/ca8542_15Jun2008_17Apr2009_082822.csav_t099.eps}{figs/ca8542_15Jun2008_17Apr2009_083623.csav_t119.eps}
 \plottwo{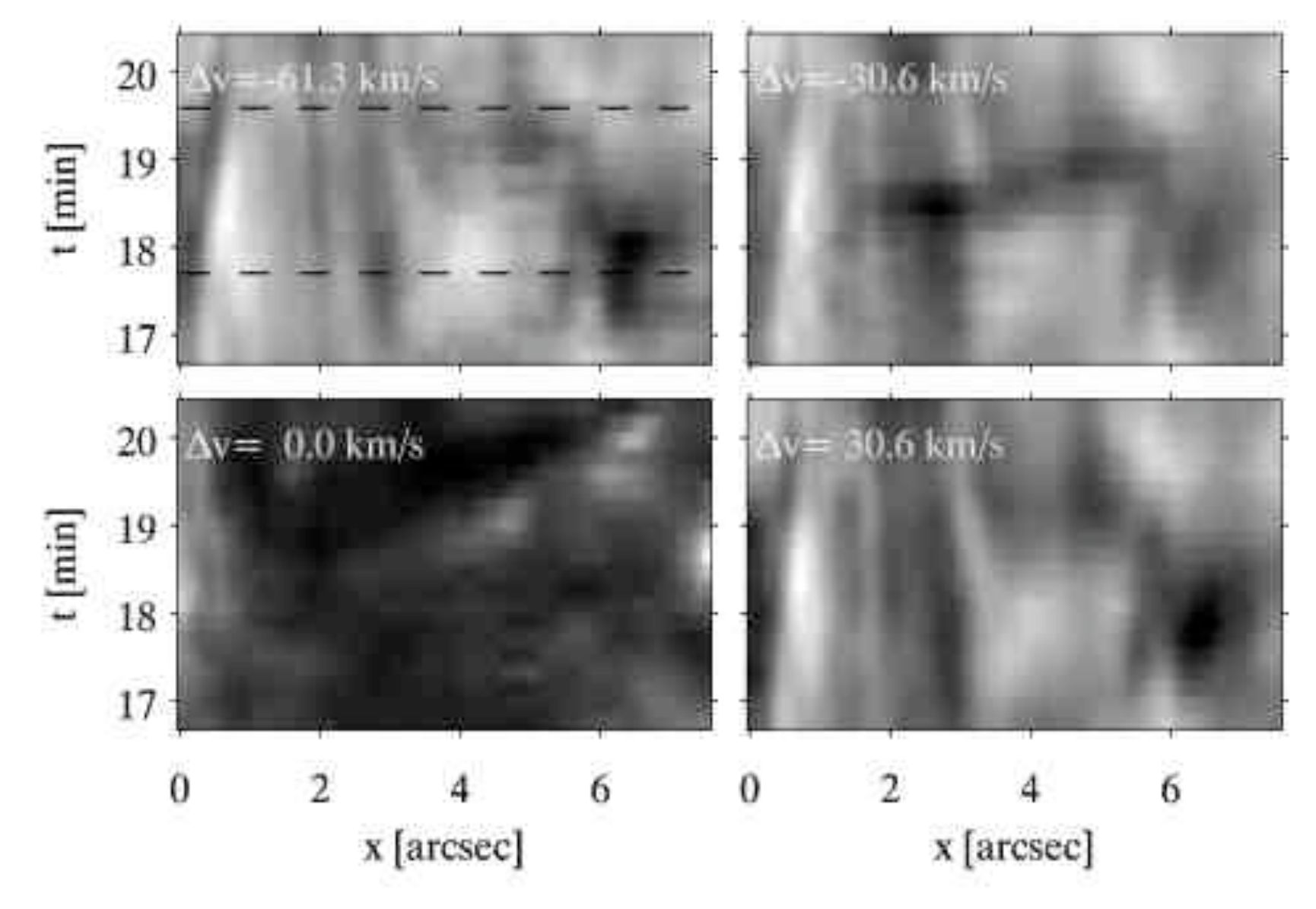}{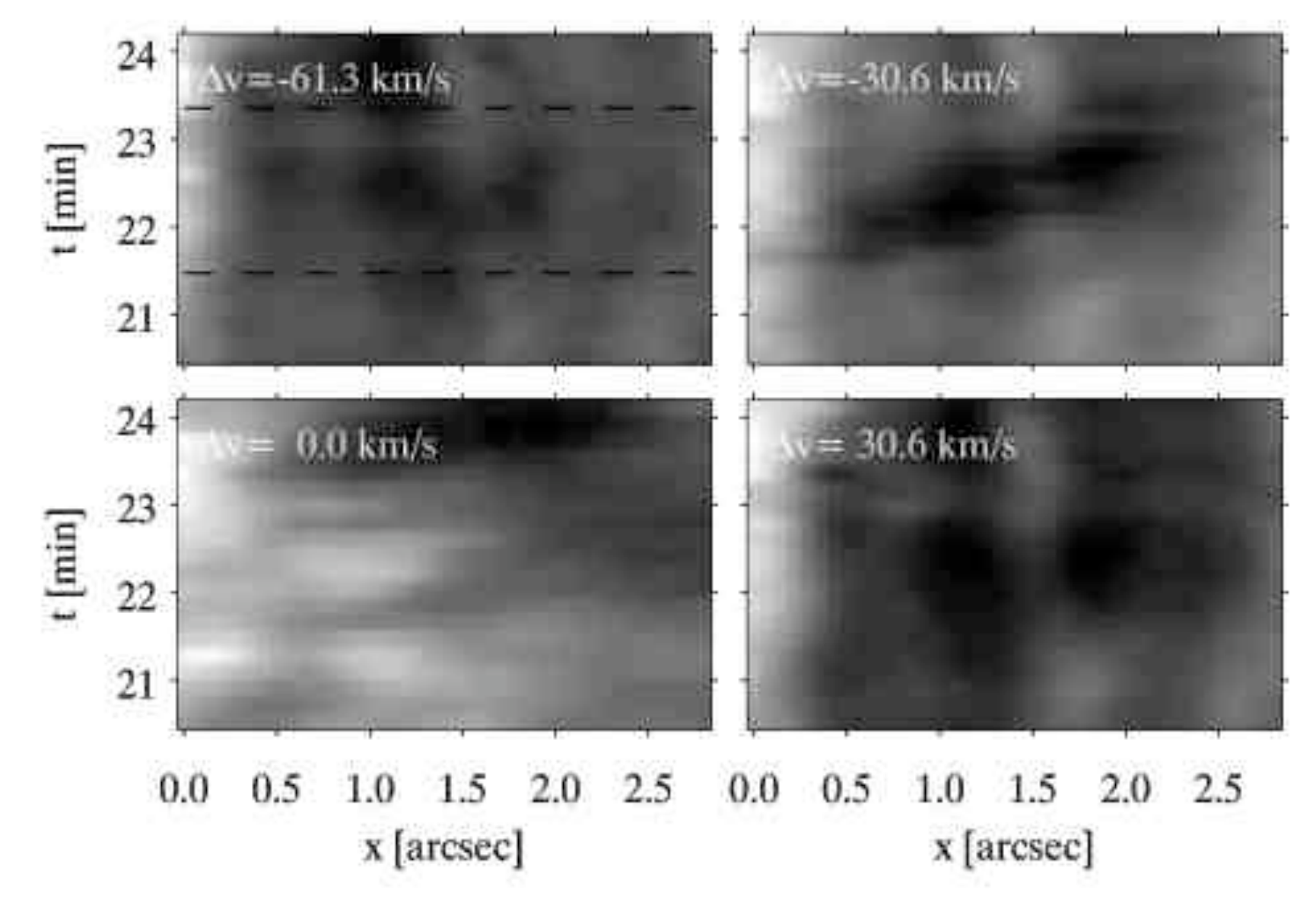}
  \caption{Timeslice along the trajectory of the RBEs in ROI4
    (left-hand panels) and ROI5 (right-hand panels). They appear as
    diagonal streaks in the $-$30.6~km~s$^{-1}$ panels (they move from left to
    right). There is no corresponding structure in the other line
    positions. The black dashed lines indicate the timespan shown in
    Fig.~\ref{fig:rbe-cuts-ca}. Note the difference in the scale along
    the x-axis between the left-hand and right-hand
    panels. \label{fig:xt-slice-ca}}
\end{figure*}

\begin{figure}
  \includegraphics[width=\columnwidth]{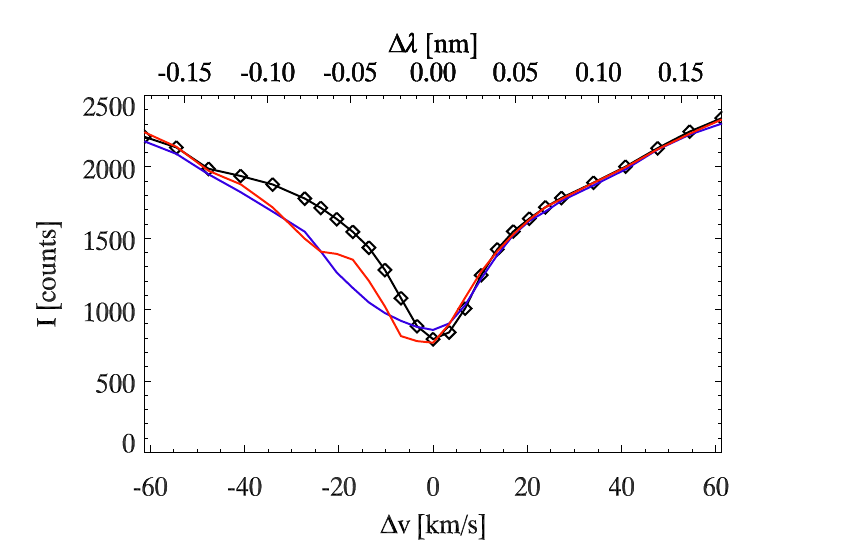}
  \caption{Example line profiles of the \CaIIIR\ RBEs. Black: average
    spectrum over the field of view, with diamonds indicating the
    different wavelength positions of the filter. Red and blue:
    profiles of RBEs. \label{fig:ca-detsp}}
\end{figure}

Figures~\ref{fig:rbe-cuts-ca}, \ref{fig:xt-slice-ca}
and~\ref{fig:ca-detsp} show details of two RBEs in the \CaIIIR\ line,
in identical format as the \Halpha\ figures. 

Figure~\ref{fig:rbe-cuts-ca} shows the image sequences of ROI4 and
ROI5. The RBEs are invisible at the largest blueshift column, instead
they appear clearly at a blueshift of $-$30.6~km~s$^{-1}$.

The left-hand panels show the appearance and disappearance of two
partially overlapping RBEs. The first RBE appears at the arrow in the
panel marked `a'. It extends, reaching maximum length in in panel b,
after which it retracts and fades. The second RBE appears in panel b,
on top of the first one, but at a slightly different angle. It gains
opacity in the following panels, bringing out its curved lower end,
indicating that it is indeed a different RBE than the first one. It
moves upward and fades from view, and is gone in the last panel. The
left-hand panels of Fig.~\ref{fig:xt-slice-ca} show timeslices of the
intensity along the trajectory of the second RBE. It shows up as a
diagonal streak in the 
%L: referee pointed at missing "-" sign
$\Delta v \is \mathrm{-}30.6$~km~s$^{-1}$
panel, again without
counterpart in the panels at other Doppler shifts.

The right-hand panels show a slow-moving roundish RBE appear at the
arrow in panel c. It moves slowly away from the network elements at
the top of the ROI and fades from view in the last two panels. The
timeslices in the right-hand panels of Fig.~\ref{fig:xt-slice-ca} show
the RBE only in the $\Delta v \is -30.6$~km~s$^{-1}$ panel.

Figure~\ref{fig:ca-detsp} shows typical line profiles of two RBEs. The
profiles are asymmetric, with extra absorption in the blue wing, and a
red wing very similar to the average profile. The largest extra
absorption occurs around $-$30 km~s$^{-1}$, at significantly lower
velocity than in the \Halpha\ profiles.

\subsection{Statistical  properties of rapid blue excursions}

\begin{figure}
  \centering
  \includegraphics[width=\columnwidth]{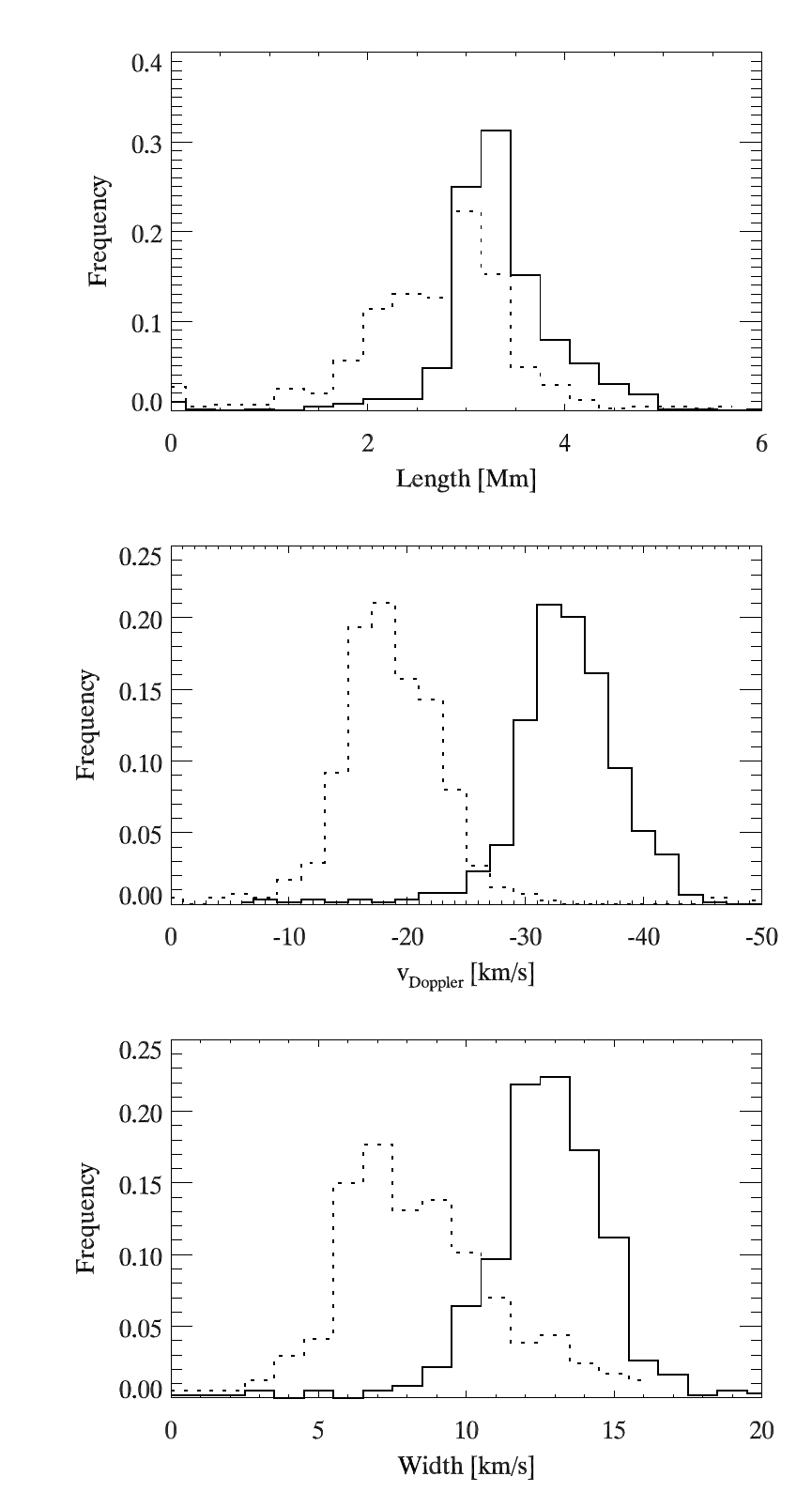}
  \caption{Histograms of RBE properties. Length (top panel), measured
    Doppler velocity at mid-point of extracted feature (middle panel) and
    width of the blue component (bottom panel) for \Halpha\ (solid line) and
    \CaIIIR\ (dashed line). \label{fig:haca_hist}}
\end{figure}

\begin{figure*}
  \centering
  \includegraphics[width=\textwidth]{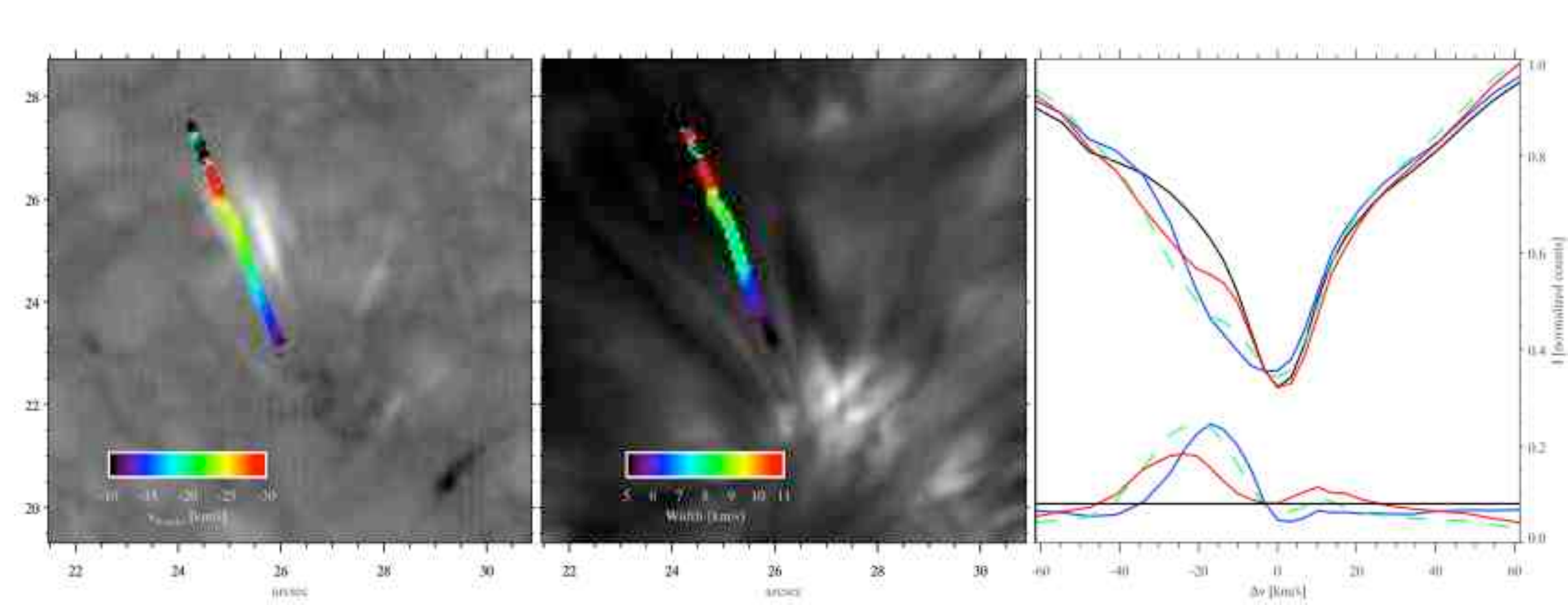}
  \includegraphics[width=\textwidth]{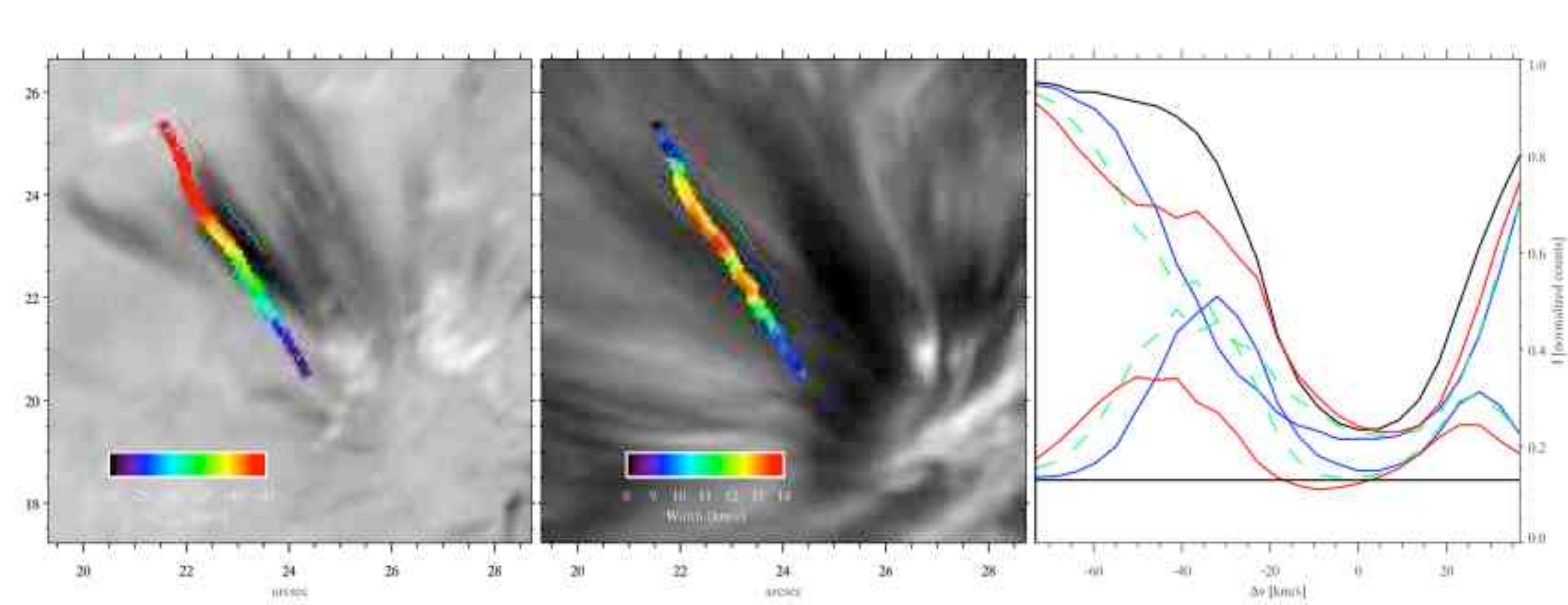}
  \caption{Properties as function of position of an RBE in the
    \CaIIIR\ dataset (top panels) and the \Halpha\ dataset (bottom
    panels). Color coded measured Doppler velocity (left panels),
    width (middle panels) and mean spectra over three portions of the
    RBE (right panels: closest to foot point in blue, middle part as
    dashed green, part furthest away in red and mean spectrum in
    black. Upper curves show the spectral profile, the lower curves show 
    the subtraction of the average spectrum and the spectral profile).  
    The extracted RBE is shown
    with a thin colored line in the left and middle panels with blue
    for the third of the length closest to the foot-point, green for
    the middle part and red for the part furthest away from the
    foot-point. The measured parameter is shown displaced to the left
    of the RBE. The background image is at line-center (middle
    panels), at a blue position of $-$36~km~s$^{-1}$ (bottom left
    panel) and a Dopplergram at 30~km~s$^{-1}$ (top left
    panel). At both the bottom and top end of the RBEs the amount of blueward 
    absorption drops significantly. When this happens we set the Doppler velocity 
    and width to zero. This is why the black color coding at both ends of the RBEs 
    is not indicative of lower velocities or widths, but rather signifies the
    spatial extent of the blue absorbing feature.
 \label{fig:haca_dop_zoom}}
\end{figure*}

\begin{figure*}
  \centering
  \includegraphics[width=\textwidth]{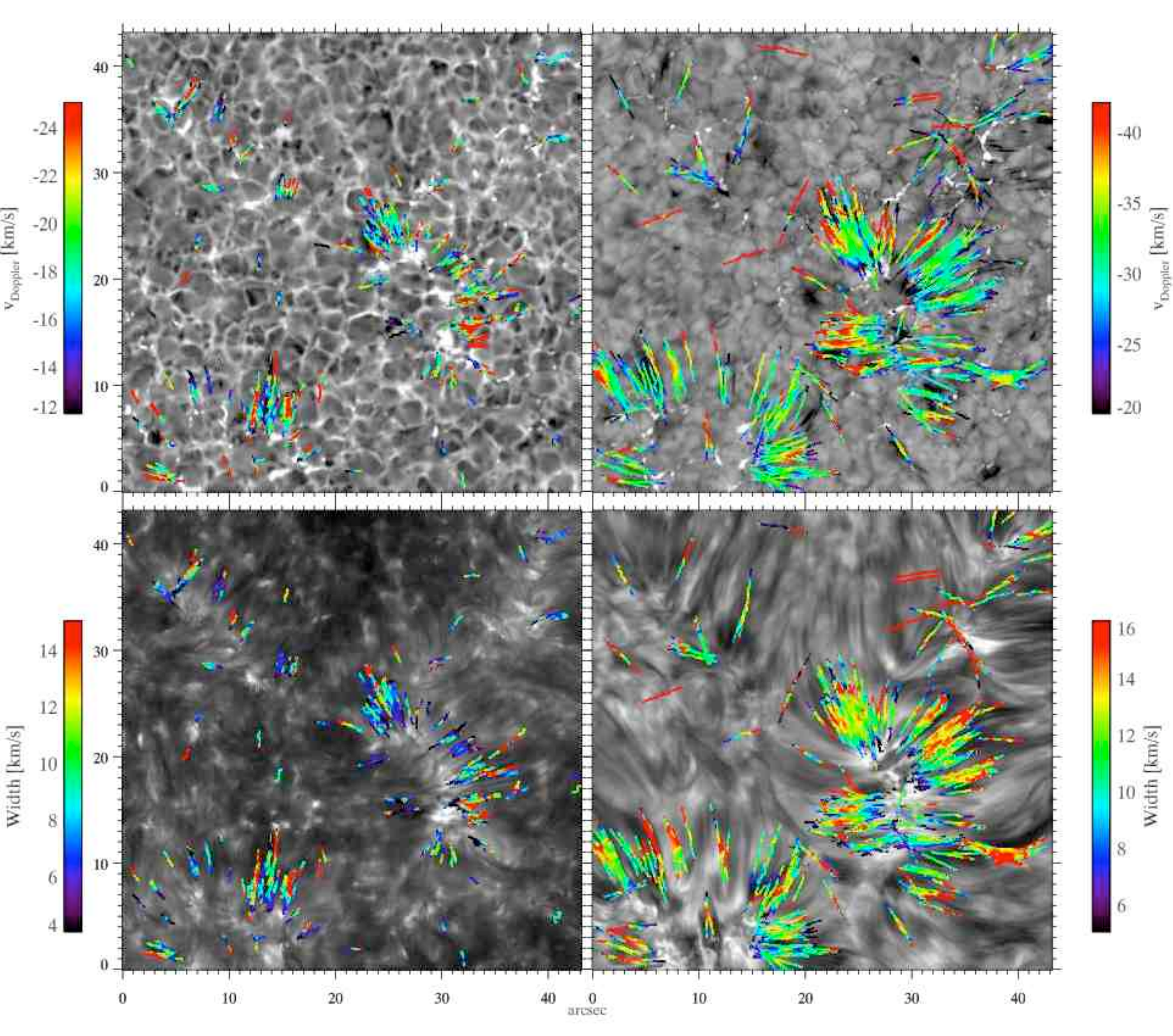}
  \caption{Properties as function of position of all detected RBEs in the \CaIIIR\ dataset
    (left panels) and the \Halpha\ dataset (right panels). Color coded measured
    Doppler velocity (top panels) and width (bottom panels) with a background image
    from $\Delta v \is -27$ km~s$^{-1}$ (top left), $\Delta v \is -46$ km~s$^{-1}$
    (top right) and line center (bottom panels).
   \label{fig:haca_vw}}
\end{figure*}

\begin{figure}
  \centering
  \includegraphics[width=\columnwidth]{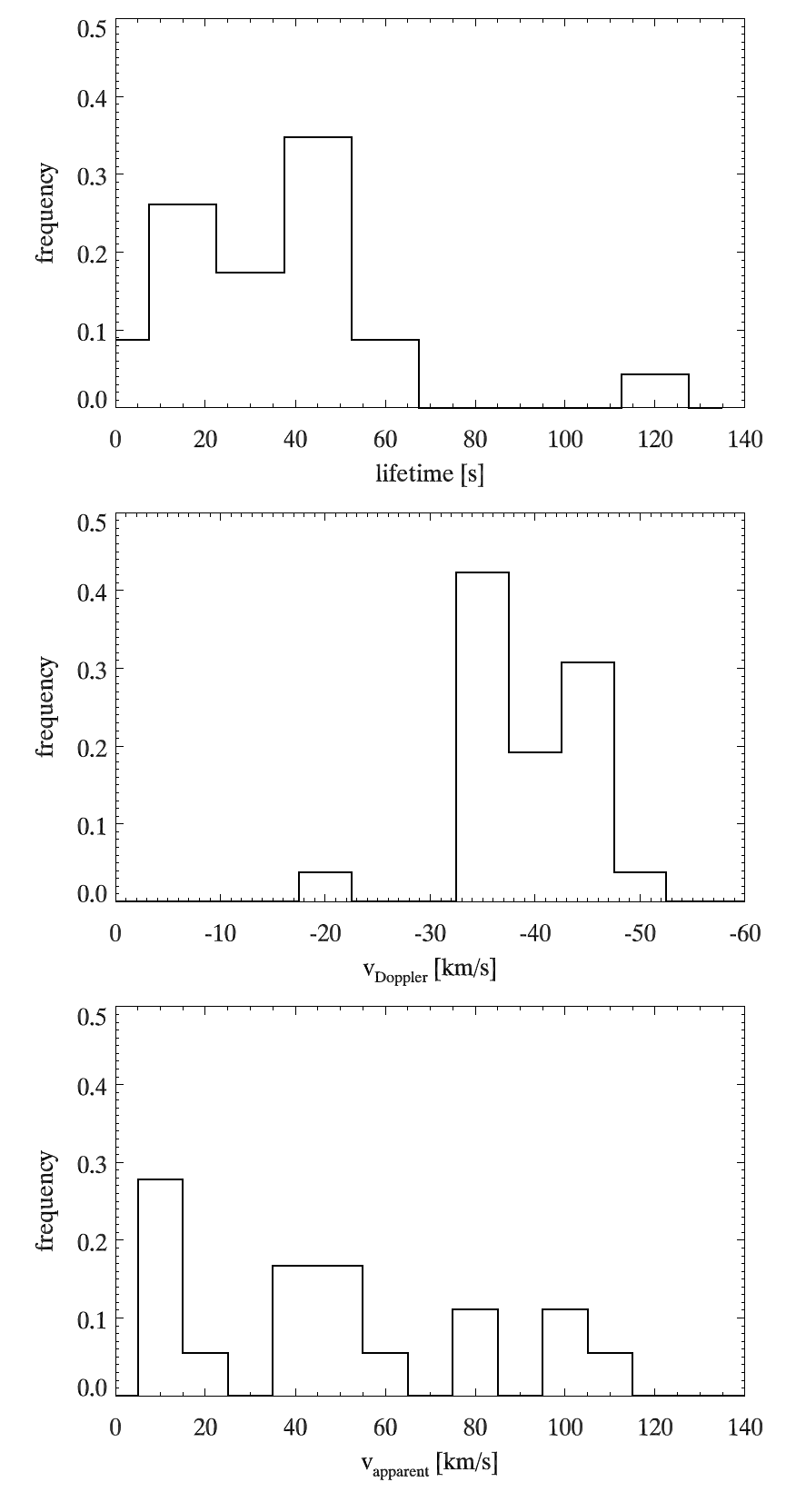}
  \caption{Histogram of RBE properties for the \Halpha\
    subsample. Lifetime (top panel), measured 
    Doppler velocity at mid-point of extracted feature (middle panel) and
    apparent velocity along the length of the RBE (bottom panel).
    \label{fig:ha_hist}}
\end{figure}

\begin{figure}
  \centering
  \includegraphics[width=\columnwidth]{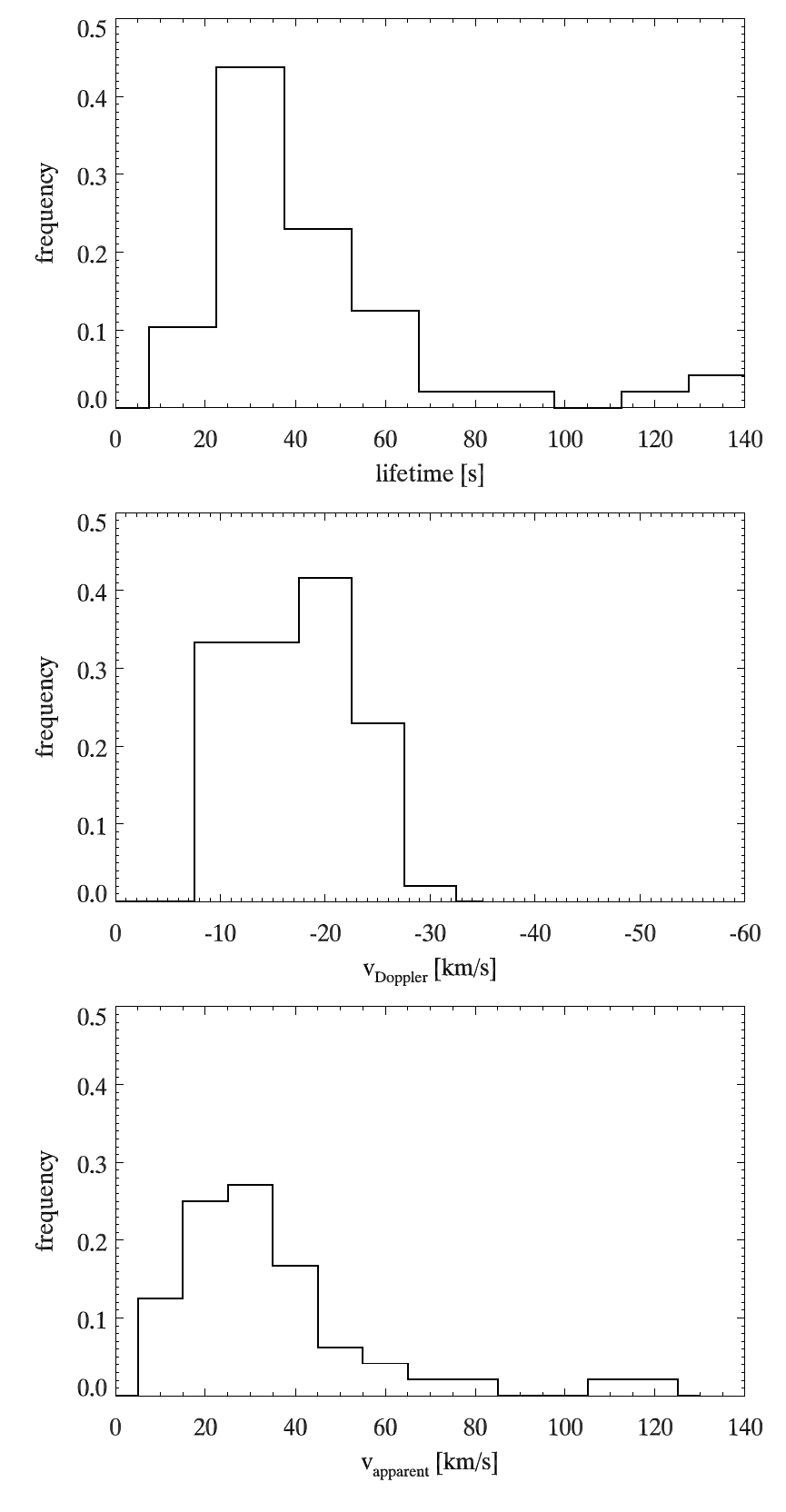}
  \caption{Histogram of RBE properties for the \CaIIIR\ subsample. Lifetime (top panel), measured
    Doppler velocity at mid-point of extracted feature (middle panel) and
    apparent velocity along the length of the RBE (bottom panel).
    \label{fig:ca_hist}}
\end{figure}

\begin{figure}
  \centering
  \includegraphics[width=\columnwidth]{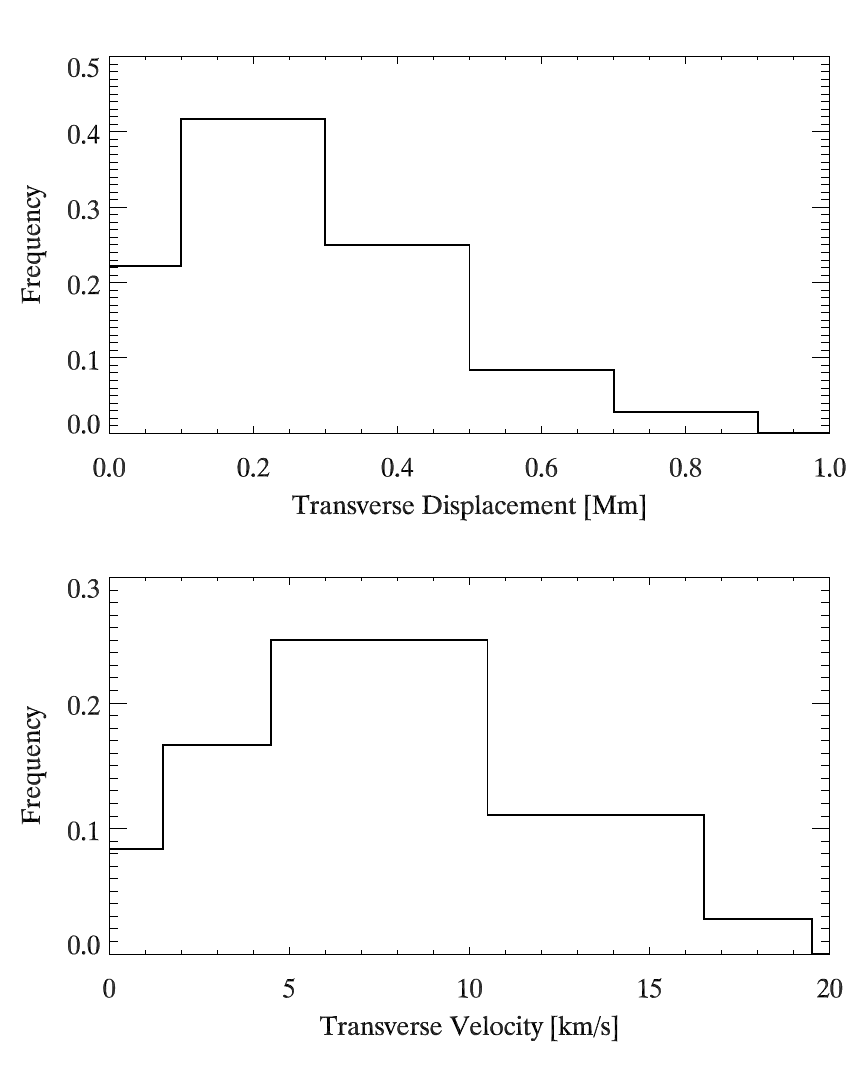}
  \caption{Histogram of transverse motions of RBEs. Transverse displacement (top panel) 
          and transverse velocity (bottom panel) of 35 \Halpha\ RBEs. Most RBEs undergo significant
          motion perpendicular to their long axis. \label{fig:haca_alfven}}
\end{figure}

The automatic detection procedure identified 413 features in the
\CaIIIR\ dataset and 608 in the \Halpha\ dataset. For each of the
identified features we subtract the average spectral profile 
(averaged over the whole field of view) from the profile at each pixel 
along the feature. We use the first and second moments with respect to
 wavelength as estimates of the Doppler velocity, $v_{\mathrm{Doppler}}$ and
 width of the blue-shifted component, $W$, i.e.,
\begin{equation}
  v_{\mathrm{Doppler}} = \frac{c}{\lambda_0} \frac{\int^{\lambda_0}_{\lambda_\mathrm{min}} (\lambda-\lambda_0)
    (I_\mathrm{avg}-I) \, \mathrm{d} \lambda}{\int^{\lambda_0}_{\lambda_\mathrm{min}}
    (I_\mathrm{avg}-I) \, \mathrm{d} \lambda},
\end{equation}
\begin{equation}
  W = \frac{c}{\lambda_0}
  \sqrt{\frac{\int^{\lambda_0}_{\lambda_\mathrm{min}} (\lambda - \lambda_{\mathrm{c}})^2
    (I_\mathrm{avg}-I) \, \mathrm{d} \lambda}{\int^{\lambda_0}_{\lambda_\mathrm{min}}
    (I_\mathrm{avg}-I) \, \mathrm{d} \lambda}},
\end{equation}
where $c$ is the velocity of light, $\lambda_0$ the line center
wavelength, 
$I$ the observed intensity, $I_{\mathrm{avg}}$ the average intensity
and $\lambda_\mathrm{c} = \lambda_0+\lambda_0 v_{\mathrm{Doppler}} / c$. The
integration ranges from the most blue-ward wavelength we observed
($\lambda_\mathrm{min}$) to line center, and is only executed for wavelengths 
for which $I_\mathrm{avg}-I > 0$, i.e., where the feature shows absorption in 
the blue wing.

We note that these estimates are necessarily rather crude, as it is
impossible to rigorously derive the atmospheric velocity structure
from the line profile. Cloud modelling might in some cases have
yielded a more precise estimate, but in the majority of cases the RBEs
are overlapping a myriad of chromospheric structures, on which
single-component cloud-model inversion is unlikely to yield reliable
results. Based on the detailed line profiles we expect our estimates
of $v_{\mathrm{Doppler}}$ to be on the low side, and most RBEs might well
have components with a $v_{\mathrm{Doppler}}$ that is 10--20 km s$^{-1}$
higher than estimated.

The measured lengths, Doppler velocities and widths are shown in
Fig.~\ref{fig:haca_hist} for both datasets. The distribution of lengths 
has a sharp lower cutoff at 0.8 Mm because this was used as a criterion for
acceptance of a feature in the automated detection algorithm. The length has
been defined as the maximum distance along which a feature shows absorption 
in the blue wing compared to the average spectral profile. This definition 
removes some of the arbitrariness introduced by the fact that the features 
are detected at a specific velocity. The values for the Doppler velocities 
and widths are calculated by averaging over the whole length of each feature.

We find that the values for length, Doppler velocity and width are consistently 
higher for the \Halpha\ RBEs compared to the \CaIIIR\ RBEs. \Halpha\ RBEs are on average
somewhat longer (of order 3 Mm vs. 2 Mm), have higher average Doppler shifts towards the blue 
(35 km s$^{-1}$ vs. 15 km s$^{-1}$) and higher widths (13 km s$^{-1}$ vs. 7 km s$^{-1}$). 
We should note that there is a wide
range of values for Doppler shifts and widths from RBE to RBE. 

In addition, the measured Doppler velocities and widths of the blue components
vary systematically along the length of the RBEs. This is shown for
two sample RBEs in Fig.~\ref{fig:haca_dop_zoom}. We find that many RBEs show an
increase of Doppler shifts and widths from their footpoint to the top. The footpoints (blue-coded
profiles and lines) typically show absorption that is at smaller Doppler shifts and narrower than
the middle (green coded) and top end (red coded) of the RBEs. We see this effect for both \Halpha\ and
\CaIIIR\ RBEs. 
This effect is visually quite striking when we plot the measured
parameters for all automatically detected RBEs for a subset of the
field of view (Fig.~\ref{fig:haca_vw}). Here we also see that the
\CaIIIR\ RBEs are typically shorter and closer to the footpoints
(photospheric magnetic field concentrations) than the \Halpha\ RBEs.

We see clear outward motions in many of the detected features but
measuring these apparent velocities is very difficult to do in an
automated fashion. Instead, we used a manual approach in which we used
the extracted positions of the features to extract $xt$-slices from
the data cube (see Fig.~\ref{fig:fov} for outlines of these
positions).
We then measured the apparent velocities,\vhor\, from the
slopes of the RBEs in these $xt$-slices manually. This was difficult
in most cases because of short lifetimes and unclear paths in the
$xt$-slices. We thus have \vhor\ for a subset of only
25 RBEs for the \Halpha\ dataset and 28 for the \CaIIIR\
dataset. There is likely to be a strong selection effect for these
measurements favoring long lifetimes and low velocities. The
lifetimes of the RBEs were determined from the first and last
occurrences in the time sequence.  The statistics for these subsamples
are shown in Figs.~\ref{fig:ha_hist} and \ref{fig:ca_hist}.
%
%The lifetimes of the RBEs range from around
%10~s (one or two successive images in our datasets), up to 120~s, with
%peaks around 40~s.
%
The lifetimes of the RBEs range from 13~s (i.e., two time steps for
\Halpha), up to 120~s, with peaks around 40~s.

The horizontal velocity of the \Halpha\ RBEs ranges from 0~km~s$^{-1}$
to 120~km~s$^{-1}$. The \CaIIIR\ RBEs have a lower maximum horizontal
velocity, with most events having $\vhor < 40$~km~s$^{-1}$.

We have also measured the transverse motions of about 35 \Halpha\ RBEs
and found that most RBEs move transversely to their long axis, with
transverse displacements of order 0.3 Mm (within a range of 0 to 0.8
Mm) and transverse velocities of order $8$ km~s$^{-1}$ (with a range
from 0 to 20 km~s$^{-1}$). This is illustrated in
Fig.~\ref{fig:haca_alfven}.  The measured lifetimes of these RBEs at
one single height are between 15 and 60 s, with an average of 33 s.

\section{Discussion \& Conclusions}

We have analyzed rapid blueshifted events in \CaIIIR\ and \Halpha\
data obtained with the CRISP instrument at the Swedish 1-m Solar
Telescope.
In blue wing time sequences, we observe these RBEs as short-lived,
narrow streaks that move at high speed away from areas with magnetic
field concentrations.
Close to disk center, the \CaIIIR\ events we observe are similar to
those found by
\citet{2008ApJ...679L.167L}. % langangen et al IBIS RBEs
We measure similar lifetimes (45~s) and Doppler velocities
(20~km~s$^{-1}$). 
Our observations have a higher spatial and temporal resolution,
allowing us to study their evolution over time and measure their
apparent motion. In addition, our automated detection algorithm allows
us to greatly expand on the statistics of RBEs.

The RBEs we identified in \Halpha\ have generally higher Doppler
velocities and larger Doppler width than their \CaIIIR\ counterparts.
The RBEs are longer in \Halpha\ than in \CaIIIR, with the \CaIIIR\
RBEs located a bit closer to the magnetic field concentrations that
are at the root of the RBEs.
For those selected RBEs where we could make a reliable estimate of the
apparent velocity, we see a slight trend of higher apparent motion in
\Halpha. The lifetimes of RBEs in \CaIIIR\ and \Halpha\ seem to be
similar.

\citet{2008ApJ...679L.167L} % langangen et al IBIS RBEs
suggested that the \CaIIIR\ RBEs are linked to the Hinode \CaIIH\ type
II spicules observed at the limb 
\citep{2007PASJ...59S.655D}.  %de pontieu et al, a tale of 2 spicules
This suggestion was based on the similarity in lifetimes, spatial
extent, location near the network and sudden disappearance or
fading. In addition, the fact that RBEs exclusively show blueshifts
corresponds well with the fact that type~II spicules only show upward
flows.

Our observations make the connection between RBEs observed on disk and
type II spicules at the limb much stronger. 
First of all we find the same similarities in lifetimes, location and
temporal evolution.  In time sequences at high Doppler displacement
from the line center, we also see the RBEs moving away from the
network as narrow streaks -- dynamical behavior that agrees well with
the upward moving type II spicules. We find that most RBEs undergo a
significant sideways motion during their lifetime. This is similar to
what \citet{2007Sci...318.1574D} observed in type~II spicules.  The
amplitude of the transverse velocities of RBEs ($\sim 8$ km~s$^{-1}$)
is slightly lower than that of type~II spicules (of order 12
km~s$^{-1}$).
Our observations show a clearly separated blueshifted component in our
line profiles which for the \Halpha\ line occurs at much higher
velocities (of order 40--50 km~s$^{-1}$) than in the \CaIIIR\
line. Such high velocities compare well with those reported for
type~II spicules (40--100 km~s$^{-1}$), especially when taking into
account the inevitable projection effects that reduce the line of
sight velocity when observing at disk center.

The differences in RBE properties between \CaIIIR\ and \Halpha\ can be
explained by the difference in chromospheric opacity of both lines,
with \Halpha\ sampling higher layers than \CaIIIR. Our finding of
higher velocities (both apparent and in Doppler) in \Halpha\ as
compared to \CaIIIR\ is highly compatible with the idea of RBEs being
disk counterparts of type II spicules: the higher opacity in \Halpha\
allows to sample higher layers where lower density plasma is propelled
to higher velocities. The fact that on average the \CaIIIR\ RBEs occur
closer to the footpoints and show lower velocities than \Halpha\ RBEs
fits in well with the Hinode/SOT limb observations which have shown
evidence of acceleration of plasma along type~II spicules
\citep{2007PASJ...59S.655D}, with higher velocities at the top of the
spicules. In addition, such acceleration is directly seen in our
individual RBEs, with increasing Doppler shifts towards the top end of
the RBEs in both \Halpha\ and \CaIIIR.

In addition to the opacity and acceleration argument, the difference
in velocities in \CaIIIR\ and \Halpha\ could also partly be explained
by the idea that
\citet{2008ApJ...679L.167L} % langangen et al IBIS RBEs
proposed. They performed numerical experiments to explore a scenario
of reconnection jets driven by constant energy release at different
atmospheric heights.
In this scenario, reconnection at lower heights (i.e., higher density)
gives rise to lower velocities -- this could be another reason why
disk center \CaIIIR\ RBEs have lower Doppler velocity than the
off-limb type II spicules.
In \Halpha\ however, higher atmospheric layers can be probed where
reconnection at lower density would lead to higher velocities.

More generally, we know that the limb observations introduce an
observational bias towards those events that rocket high above the
chromospheric fibrilar junk at lower heights.
The large line of sight obscuration close to the limb may well, to a
large extent, hide the lower velocity type II events at lower
inclination with the surface.
The highest velocities ($\sim 100$ km s$^{-1}$) of the type~II
spicules may thus be missed on the disk because of this selection
effect, in addition to the low opacity of the low density tops of the
spicules. The latter effect may also be the reason for the slightly
lower transverse velocities we find in the \Halpha\ RBEs compared to
those of the type~II spicules at the limb. Generally, the RBEs likely
sample slightly lower regions in the Ca~II spicules where both
longitudinal and transverse velocities are also a bit lower.
 
These and other selection effects imply that the histograms of
Figs.~\ref{fig:ha_hist} and~\ref{fig:ca_hist} should not be interpreted
as ``the'' properties of RBEs. After all, it is quite possible that
our selection criteria might exclude events that have too little
absorption or absorption at different Doppler velocities. In addition,
the velocities we measure here are only rough estimates of the real
mass motion. For example, the apparent velocities can sometimes be a
bit more difficult to measure and interpret.  This is because visual
inspection of the movies at various wavelengths suggests that
measurements of such projected velocities can sometimes depend
considerably on the wavelength studied. However, the presence of a
separate blue-shifted absorbing component certainly indicates strong
mass flows of at least 30--50 km~s$^{-1}$ in these features.

The events we measure thus have a wide variety of properties
(lifetimes, velocities, temporal evolution, location, acceleration,
etc.) that are highly similar to those of the type~II spicules
observed in \CaIIH\ on the limb. We are therefore confident that we
have found the on-disk counterparts of type II spicules.

This finding is further strengthened when we consider the occurrence
rate of RBEs and compare it to the occurrence rate of type~II spicules
at the limb. The latter can be deduced from a visual analysis of
Hinode/SOT data such as that studied by
\citet{2007PASJ...59S.655D}. There we find between 1.5 and 3 \CaIIH\
type~II spicules per linear arcsec along the limb in a coronal
hole. To compare this to our disk center observations of a coronal
hole, we use our automated algorithm on the \Halpha\ images at $-$35
km~s$^{-1}$ (the peak of the RBE Doppler shift distribution,
Fig.~\ref{fig:haca_hist}) and find on average about 40 RBEs per
image. The field of view is about 70\arcsec by 70\arcsec.
Observations at the limb show a large amount of line of sight
superposition since spicules are quite tall (up to 10,000 km): we can
still discern spicules (of 5,000 km height) at the limb even when they
occur up to 115 arcsec in front or behind the limb. This means that a
typical limb observation samples spicules over a spatial range of 230
arcseconds along the line of sight. This implies that our 40 RBEs per
image would translate to $40\times230/70=131$ RBEs per 70 arcsec along
the limb, i.e., 1.9 RBEs per linear arcsec.  This compares very
favorably with the 1.5 to 3 type~II spicules we observe on average in
Hinode/SOT limb data.

The discovery of the disk counterpart of type~II spicules provides
exciting new avenues towards resolving several of the major unresolved
issues of these features. By providing a top view that is unhampered
by the enormous line of sight superposition at the limb, RBEs will
allow for a much improved study of the formation mechanism that drives
these highly dynamic jets. Our preliminary finding that many of the
RBEs show an increase in velocity and width of the blueshifted
absorbing component as one travels along the jet provides a much
needed constraint on theoretical models. For example, such an increase
in velocity along the jet may well be compatible with (instantaneous)
velocity profiles along reconnection jets in 2D MHD simulations
\citep[see, e.g.][]{2009arXiv0902.0977H}.  The increase in width of
the absorbing component along the feature, and the sudden
disappearance or fading of RBEs at the end of their lifetime are
highly compatible with a scenario that includes ongoing heating of the
plasma along the jet. Such heating provides strict constraints on the
acceleration and heating process that drives these jets. Detailed
comparisons of the statistical properties of RBEs with advanced
numerical models will be required to ascertain whether reconnection or
other formation mechanisms (e.g., energy deposition at chromospheric
heights from electron beams) can produce the details of the observed
velocity and width profiles.

Now that we have established the disk counterpart of type~II spicules,
we will also be able to study in detail the relationship between these
jets and the magnetic field concentrations with which they are
associated. Such a relationship is bound to be quite complex, since we
observe RBEs not only in association with mostly unipolar network in
coronal holes, but also in quiet Sun, and highly unipolar plage
regions.  If reconnection plays a role in the formation of these jets,
it is likely that component reconnection (e.g., at tangential
discontinuities) plays a significant role
\citep[see e.g.,][]{parker94current_sheets}. % Spontaneous current
                                % sheets in magnetic fields 
By studying the underlying
magnetic field concentrations and its interaction with the flowfield,
as well as the spectral line profiles at the footpoints of the RBEs,
we should be able to constrain the various suggested driving
mechanisms.

This is exemplified by the discovery of the black beads. These
roundish features with strong absorption in the blue wing have many
properties of RBEs with some significant differences. They appear
along a neutral line between two regions of opposite polarity, and not
around unipolar network like the other RBEs. Their shape and high
velocities suggests that these beads may be jets in the direction of
the line of sight. This can help explain the extremely high Doppler
velocities in excess of $-$74~km~s$^{-1}$ in the beads. It is tempting
to speculate that the different magnetic topology also plays a role in
the larger velocities. Perhaps the reconnection events release more
energy because of opposite-polarity cancellation instead of
reconnection at a tangential discontinuity? A preliminary study of the
line profiles of the black beads indicates that some profiles are
broadened on the red side of the spectral line as well. This could be
indicative of heating at the footpoint of the jet. We will study these
events and more generally the spectral line profiles at the footpoints
of RBEs in a more extensive follow-up paper.

The presence of RBEs on the solar disk also provides an exciting way
of establishing the connection of these jets to the recently
discovered upflows that are seen as faint asymmetric profiles of EUV
and UV emission lines which are formed at transition region and
coronal temperatures \citep{2009arXiv0906.5434D}. These upflows in the
transition region and corona have velocities of order 50--100
km~s$^{-1}$ and have been connected to so-called straws: jet-like
features observed as highly dynamic elongated brightenings on the disk
in the Hinode/SOT \CaIIH\ passband. Since the observations of
\citet{2009arXiv0906.5434D} lacked chromospheric velocity information,
significant uncertainties remain about the heating of type~II spicules
to coronal temperatures. The velocities, thermal evolution and
occurrence rate of the RBEs we report on here are certainly compatible
with the scenario of \citet{2009arXiv0906.5434D}. Follow-up
observational studies with Hinode and CRISP will be invaluable to
fully determine the thermal evolution of these spicules and the role
they play in providing the corona with hot plasma.

\acknowledgments
% La Palma:
This research was supported by the Research Council of Norway through
grant 170935/V30.
% Jorrit:
J.L. is supported by the European Commission funded
Research Training Network SOLAIRE. 
% Bart:
B.D.P. thanks the Oslo group for excellent hospitality and is
supported by NASA grants NNM07AA01C (HINODE), NNG06GG79G and
NNX08AH45G.
% Gregal:
G.V. is supported by a Marie Curie Early Stage Research Training
Fellowship of the European Community's 6th Framework Programme
(MEST-CT-2005-020395): The USO-SP International School for Solar
Physics.
% fellow observers and ISP staff:
We thank Ada Ortiz Carbonell, Viggo Hansteen, Sven Wedemeyer-B{\"o}hm,
Pit S{\"u}tterlin, and Michiel van Noort for their help during the
observations.
% and Viggo again:
We thank Viggo Hansteen for extensive discussions.
% SST:
The Swedish 1-m Solar Telescope is operated on the island of La Palma
by the Institute for Solar Physics of the Royal Swedish Academy of
Sciences in the Spanish Observatorio del Roque de los Muchachos of the
Instituto de Astrof{\'\i}sica de Canarias.

\bibliographystyle{aa} 
\bibliography{spics2}

\begin{thebibliography}{28}
\expandafter\ifx\csname natexlab\endcsname\relax\def\natexlab#1{#1}\fi

\bibitem[{{Beckers}(1968)}]{1968SoPh....3..367B}
{Beckers}, J.~M. 1968, \solphys, 3, 367

\bibitem[{{Chae} {et~al.}(1998){Chae}, {Wang}, {Lee}, {Goode}, \&
  {Schuhle}}]{1998ApJ...504L.123C}
{Chae}, J., {Wang}, H., {Lee}, C.-Y., {Goode}, P.~R., \& {Schuhle}, U. 1998,
  \apjl, 504, L123+

\bibitem[{{De Pontieu} {et~al.}(2007{\natexlab{a}}){De Pontieu}, {Hansteen},
  {Rouppe van der Voort}, {van Noort}, \& {Carlsson}}]{2007ApJ...655..624D}
{De Pontieu}, B., {Hansteen}, V.~H., {Rouppe van der Voort}, L., {van Noort},
  M., \& {Carlsson}, M. 2007{\natexlab{a}}, ApJ, 655, 624

\bibitem[{{De Pontieu} {et~al.}(2007{\natexlab{b}}){De Pontieu}, {McIntosh},
  {Hansteen}, {Carlsson}, {Schrijver}, {Tarbell}, {Title}, {Shine}, {Suematsu},
  {Tsuneta}, {Katsukawa}, {Ichimoto}, {Shimizu}, \&
  {Nagata}}]{2007PASJ...59S.655D}
{De Pontieu}, B., {McIntosh}, S., {Hansteen}, V.~H., {et~al.}
  2007{\natexlab{b}}, \pasj, 59, 655

\bibitem[{{De Pontieu} {et~al.}(2007{\natexlab{c}}){De Pontieu}, {McIntosh},
  {Carlsson}, {Hansteen}, {Tarbell}, {Schrijver}, {Title}, {Shine}, {Tsuneta},
  {Katsukawa}, {Ichimoto}, {Suematsu}, {Shimizu}, \&
  {Nagata}}]{2007Sci...318.1574D}
{De Pontieu}, B., {McIntosh}, S.~W., {Carlsson}, M., {et~al.}
  2007{\natexlab{c}}, Science, 318, 1574

\bibitem[{{De Pontieu} {et~al.}(2009){De Pontieu}, {McIntosh}, {Hansteen}, \&
  {Schrijver}}]{2009arXiv0906.5434D}
{De Pontieu}, B., {McIntosh}, S.~W., {Hansteen}, V.~H., \& {Schrijver}, C.~J.
  2009, ArXiv e-prints

\bibitem[{{Hansteen} {et~al.}(2006){Hansteen}, {De Pontieu}, {Rouppe van der
  Voort}, {van Noort}, \& {Carlsson}}]{2006ApJ...647L..73H}
{Hansteen}, V.~H., {De Pontieu}, B., {Rouppe van der Voort}, L., {van Noort},
  M., \& {Carlsson}, M. 2006, \apjl, 647, L73

\bibitem[{{Heggland} {et~al.}(2007){Heggland}, {De Pontieu}, \&
  {Hansteen}}]{2007ApJ...666.1277H}
{Heggland}, L., {De Pontieu}, B., \& {Hansteen}, V.~H. 2007, \apj, 666, 1277

\bibitem[{{Heggland} {et~al.}(2009){Heggland}, {De Pontieu}, \&
  {Hansteen}}]{2009arXiv0902.0977H}
{Heggland}, L., {De Pontieu}, B., \& {Hansteen}, V.~H. 2009, ArXiv e-prints

\bibitem[{{Kosugi} {et~al.}(2007){Kosugi}, {Matsuzaki}, {Sakao}, {Shimizu},
  {Sone}, {Tachikawa}, {Hashimoto}, {Minesugi}, {Ohnishi}, {Yamada}, {Tsuneta},
  {Hara}, {Ichimoto}, {Suematsu}, {Shimojo}, {Watanabe}, {Shimada}, {Davis},
  {Hill}, {Owens}, {Title}, {Culhane}, {Harra}, {Doschek}, \&
  {Golub}}]{2007SoPh..243....3K}
{Kosugi}, T., {Matsuzaki}, K., {Sakao}, T., {et~al.} 2007, \solphys, 243, 3

\bibitem[{{Langangen} {et~al.}(2008){Langangen}, {De Pontieu}, {Carlsson},
  {Hansteen}, {Cauzzi}, \& {Reardon}}]{2008ApJ...679L.167L}
{Langangen}, {\O}., {De Pontieu}, B., {Carlsson}, M., {et~al.} 2008, \apjl,
  679, L167

\bibitem[{{Leenaarts} {et~al.}(2009){Leenaarts}, {Carlsson}, {Hansteen}, \&
  {Rouppe van der Voort}}]{2009ApJ...694L.128L}
{Leenaarts}, J., {Carlsson}, M., {Hansteen}, V., \& {Rouppe van der Voort}, L.
  2009, \apjl, 694, L128

\bibitem[{{Leenaarts} {et~al.}(2006{\natexlab{a}}){Leenaarts}, {Rutten},
  {Carlsson}, \& {Uitenbroek}}]{2006A&A...452L..15L}
{Leenaarts}, J., {Rutten}, R.~J., {Carlsson}, M., \& {Uitenbroek}, H.
  2006{\natexlab{a}}, \aap, 452, L15

\bibitem[{{Leenaarts} {et~al.}(2006{\natexlab{b}}){Leenaarts}, {Rutten},
  {S{\"u}tterlin}, {Carlsson}, \& {Uitenbroek}}]{2006A&A...449.1209L}
{Leenaarts}, J., {Rutten}, R.~J., {S{\"u}tterlin}, P., {Carlsson}, M., \&
  {Uitenbroek}, H. 2006{\natexlab{b}}, \aap, 449, 1209

\bibitem[{{Martinez-Sykora} {et~al.}(2009){Martinez-Sykora}, {Hansteen}, {De
  Pontieu}, \& {Carlsson}}]{2009arXiv0906.4446M}
{Martinez-Sykora}, J., {Hansteen}, V., {De Pontieu}, B., \& {Carlsson}, M.
  2009, ArXiv e-prints

\bibitem[{{Parker}(1994)}]{parker94current_sheets}
{Parker}, E. 1994, Spontaneous current sheets in magnetic fields : with
  applications to stellar x-rays (Oxford University Press)

\bibitem[{{Rouppe van der Voort} {et~al.}(2007){Rouppe van der Voort}, {De
  Pontieu}, {Hansteen}, {Carlsson}, \& {van Noort}}]{2007ApJ...660L.169R}
{Rouppe van der Voort}, L.~H.~M., {De Pontieu}, B., {Hansteen}, V.~H.,
  {Carlsson}, M., \& {van Noort}, M. 2007, \apjl, 660, L169

\bibitem[{{Scharmer}(2006)}]{2006A&A...447.1111S}
{Scharmer}, G.~B. 2006, \aap, 447, 1111

\bibitem[{{Scharmer} {et~al.}(2003{\natexlab{a}}){Scharmer}, {Bjelksjo},
  {Korhonen}, {Lindberg}, \& {Petterson}}]{2003SPIE.4853..341S}
{Scharmer}, G.~B., {Bjelksjo}, K., {Korhonen}, T.~K., {Lindberg}, B., \&
  {Petterson}, B. 2003{\natexlab{a}}, in Society of Photo-Optical
  Instrumentation Engineers (SPIE) Conference Series, Vol. 4853, Society of
  Photo-Optical Instrumentation Engineers (SPIE) Conference Series, ed. S.~L.
  {Keil} \& S.~V. {Avakyan}, 341--350

\bibitem[{{Scharmer} {et~al.}(2003{\natexlab{b}}){Scharmer}, {Dettori},
  {Lofdahl}, \& {Shand}}]{2003SPIE.4853..370S}
{Scharmer}, G.~B., {Dettori}, P.~M., {Lofdahl}, M.~G., \& {Shand}, M.
  2003{\natexlab{b}}, in Presented at the Society of Photo-Optical
  Instrumentation Engineers (SPIE) Conference, Vol. 4853, Society of
  Photo-Optical Instrumentation Engineers (SPIE) Conference Series, ed. S.~L.
  {Keil} \& S.~V. {Avakyan}, 370--380

\bibitem[{{Scharmer} {et~al.}(2008){Scharmer}, {Narayan}, {Hillberg}, {de la
  Cruz Rodriguez}, {L{\"o}fdahl}, {Kiselman}, {S{\"u}tterlin}, {van Noort}, \&
  {Lagg}}]{2008ApJ...689L..69S}
{Scharmer}, G.~B., {Narayan}, G., {Hillberg}, T., {et~al.} 2008, \apjl, 689,
  L69

\bibitem[{{Sterling}(2000)}]{2000SoPh..196...79S}
{Sterling}, A.~C. 2000, \solphys, 196, 79

\bibitem[{{Tsuneta} {et~al.}(2008){Tsuneta}, {Ichimoto}, {Katsukawa}, {Nagata},
  {Otsubo}, {Shimizu}, {Suematsu}, {Nakagiri}, {Noguchi}, {Tarbell}, {Title},
  {Shine}, {Rosenberg}, {Hoffmann}, {Jurcevich}, {Kushner}, {Levay}, {Lites},
  {Elmore}, {Matsushita}, {Kawaguchi}, {Saito}, {Mikami}, {Hill}, \&
  {Owens}}]{2008SoPh..249..167T}
{Tsuneta}, S., {Ichimoto}, K., {Katsukawa}, Y., {et~al.} 2008, \solphys, 249,
  167

\bibitem[{{van Noort} {et~al.}(2005){van Noort}, {Rouppe van der Voort}, \&
  {L{\"o}fdahl}}]{2005SoPh..228..191V}
{van Noort}, M., {Rouppe van der Voort}, L., \& {L{\"o}fdahl}, M.~G. 2005,
  \solphys, 228, 191

\bibitem[{{van Noort} \& {Rouppe van der Voort}(2006)}]{2006ApJ...648L..67V}
{van Noort}, M.~J. \& {Rouppe van der Voort}, L.~H.~M. 2006, ApJL, 648, L67

\bibitem[{{van Noort} \& {Rouppe van der Voort}(2008)}]{2008A&A...489..429V}
{van Noort}, M.~J. \& {Rouppe van der Voort}, L.~H.~M. 2008, \aap, 489, 429

\bibitem[{{Vecchio} {et~al.}(2009){Vecchio}, {Cauzzi}, \&
  {Reardon}}]{2009A&A...494..269V}
{Vecchio}, A., {Cauzzi}, G., \& {Reardon}, K.~P. 2009, \aap, 494, 269

\bibitem[{{Wang} {et~al.}(1998){Wang}, {Chae}, {Gurman}, \&
  {Kucera}}]{1998SoPh..183...91W}
{Wang}, H., {Chae}, J., {Gurman}, J.~B., \& {Kucera}, T.~A. 1998, \solphys,
  183, 91

\end{thebibliography}

%\begin{thebibliography}{}
%\end{thebibliography}

\end{document}